\documentclass[twocolumn,aps,nofootinbib]{revtex4-1}
\usepackage{amsfonts, amssymb}
\usepackage{graphicx, epsfig,bm}
\usepackage{color}

\usepackage{amsbsy}
\usepackage{amsfonts}
\usepackage{amsmath}
\usepackage{amssymb}

\newcommand{\be}{\begin{equation}}
\newcommand{\ee}{\end{equation}}
\newcommand{\bea}{\begin{eqnarray}}
\newcommand{\eea}{\end{eqnarray}}

\def\fun#1#2{\lower3.6pt\vbox{\baselineskip0pt\lineskip.9pt
        \ialign{$\mathsurround=0pt#1\hfill##\hfil$\crcr#2\crcr\sim\crcr}}}

\renewcommand\){\right)}
\renewcommand\[{\left[}
\renewcommand\]{\right]}
\newcommand{\el}{l}
\newcommand{\wj}[6]{\left(
                           \begin{array}{ccc}
        \! #1\! & #2\!  & #3\!  \\
        \! #4\! & #5\!  & #6\!
                           \end{array}
                   \right)}

\newcommand\lsim{\mathrel{\rlap{\lower4pt\hbox{\hskip1pt$\sim$}}
    \raise1pt\hbox{$<$}}}
\newcommand\gsim{\mathrel{\rlap{\lower4pt\hbox{\hskip1pt$\sim$}}
    \raise1pt\hbox{$>$}}}

\def\dslash{\not{\hbox{\kern-2pt $\partial$}}}
\def\Dslash{\not{\hbox{\kern-4pt $D$}}}
\def\Oslash{\not{\hbox{\kern-4pt $O$}}}
\def\Qslash{\not{\hbox{\kern-4pt $Q$}}}
\def\pslash{\not{\hbox{\kern-2.3pt $p$}}}
\def\kslash{\not{\hbox{\kern-2.3pt $k$}}}
\def\qslash{\not{\hbox{\kern-2.3pt $q$}}}

 \newtoks\slashfraction
 \slashfraction={.13}
 \def\slash#1{\setbox0\hbox{$ #1 $}
 \setbox0\hbox to \the\slashfraction\wd0{\hss \box0}/\box0 }

\def\ee{\end{equation}}
\def\be{\begin{equation}}

\newcommand\Tr{{\rm Tr}\,}

\begin{document}
\setlength{\unitlength}{1mm}
\title{Dark Energy and Neutrino Masses from Future Measurements
  of the \\Expansion History and Growth of Structure}  
\author{Shahab Joudaki, Manoj Kaplinghat}
\affiliation{Center for Cosmology, Dept. of Physics \& Astronomy, University of California, Irvine, CA 92697}

\date{\today}

\begin{abstract}
We forecast the expected cosmological constraints from a combination of
probes of both the universal expansion rate and matter perturbation
growth, in the form of weak lensing tomography, galaxy tomography,
supernovae, and the cosmic microwave background  
incorporating all cross-correlations between the observables for an
extensive cosmological parameter set.  
We allow for non-zero curvature and parameterize our ignorance of
the early universe by allowing for a 
non-negligible fraction of dark energy (DE) at high redshifts. We find
that early DE density can be constrained to 0.2\% of the critical
density of the universe with Planck combined with a ground-based
LSST-like survey, while curvature can be constrained to
0.06\%. However, these additional degrees of freedom degrade  
our ability to measure late-time dark energy and the sum of neutrino
masses. We find that the combination of cosmological
probes can break degeneracies and constrain the sum of neutrino masses
to 0.04 eV,  present DE density also to 0.2\% of the critical density,
and the equation of state to 0.01 -- roughly a factor of two
degradation in the constraints overall compared to the case without
allowing for early DE.
The constraints for a space-based mission are similar. Even a modest
1\% dark energy fraction of the critical density at high redshift, if
not accounted for in future analyses, biases the cosmological
parameters by up to $2\sigma$. Our analysis suggests that throwing out
nonlinear scales (multipoles $> 1000$) may not result in significant
degradation in future parameter measurements when multiple
cosmological probes are combined. We find that including
cross-correlations between the different probes can result in
improved constraints by up to a factor of 2 for the sum of
neutrino masses and early dark energy density.
\end{abstract}
\bigskip

\maketitle

\section{Introduction}
The measurement of the low-redshift expansion history through
observations of SNIa \cite{RiessSNe,Perlmutter}, combined with cosmic
microwave background (CMB) observations of flatness \cite{Dunkley,Komatsu} and 
large-scale structure (LSS) measurements of a sub-critical matter 
density~\cite{Tegmark:2003uf,Tegmark:2006az,Cole:2005sx} provide 
strong arguments  for a currently accelerating phase in the expansion of the universe. 
The simplest solution to these  observations is provided by a cosmological constant, or a 
uniformly smooth vacuum energy, with a pressure that is the negative of the energy density. 

Among alternatives to the cosmological constant, the most popular are
scalar field models with potentials tailored to give rise to late-time
acceleration and current equation of state for the dark energy, $w$, close
to -1~\cite{Ford:1987de, Ratra:1987rm, Wetterich88, Peebles:1987ek, Zlatev, Ferreira, Caldwell:1997ii, Chiba}.
These models are fine-tuned to have dark energy dominate today, just
like the cosmological constant. However, the requirement $w \gtrsim
-1$ currently, does not imply that dark energy was sub-dominant at
earlier times, specifically redshift $z \gtrsim 2$, where we have no
direct constraints. Even in single scalar field models, one could have
a wide range of behavior for $w(z)$, or equivalently the evolution of
the dark energy density with redshift. For example, oscillating $w(z)$
models provide an example where the dark energy
density is not negligible in the past, while at the same time the
fine-tuning of the potential is benign as compared to anthropic considerations~\cite{Dodelson:2001fq}.

Observationally, we have no direct constraints on the expansion rate
of the universe at $z \gtrsim 2$. One way to parameterize our
ignorance is to allow for early dark energy. Such a parameterization
also allows us to estimate the changes in the growth of structure
compared to a $\Lambda$CDM model.

Our aim in this paper is three-fold.
\begin{enumerate}
\item We explore how well the present dark energy density and its
  equation of state may be constrained using multiple probes that are
  sensitive to the growth of structure and expansion history. {\em We 
  include  cross-correlations between these different probes and
  isolate their effect on parameter constraints and degeneracies.}  
\item We include non-negligible dark energy density at early times
  ($z\gtrsim 2$), ask how well this contribution can be measured, and
  how it affects the inferred values for the present dark energy
  equation of state $w$ and density $\Omega_{d0}$. 
\item We discuss how the inclusion of early dark energy affects the
  determination of the sum of the neutrino masses in spatially flat
  and non-flat cosmological models, along with the expected
  degradation from unknown reionization history. 
\end{enumerate}
With respect to the second point above, we note that a study of
expected constraints from baryon acoustic oscillation (BAO)
observations has shown that the presence of early dark 
energy can significantly bias the inferred values of late-time dark 
energy parameters~\cite{LinRob}.
We may attempt to resolve this bias by using a calibration parameter,
similar to that for SNe observations. However, this would lead to more
than a factor of two larger degradation in BAO dark energy
constraints~\cite{LinRob}, which stresses the importance of including
this uncertainty in parameter constraints and on measuring dark energy
contribution to the expansion rate at $z \gtrsim 2$.  

At low redshift, the expansion rate in a model with EDE is designed to
masquerade that of models with late-time ($z \lesssim 1$) dark
energy~\cite{DorRob}, thereby severely limiting the constraining
ability of probes exclusively sensitive to the universal expansion,
such as SN measurements. The handle on early dark energy must
therefore come from a combination of expansion rate and matter
power spectrum measurements.   

For a universe with dark energy to produce the matter perturbations
seen in the present universe, more structure needed to have formed at
earlier times than for a universe without dark energy. Similarly, a
universe with a non-vanishing amount of dark energy at early times
requires more structure to have formed at earlier times than a
universe with only late-time dark energy. With a fixed large-scale
normalization of the power spectrum, this implies a suppression of the
matter power spectrum on small scales. 

Another effect on small scales is due to massive neutrinos.These
relativistic massive neutrinos can stream out of high-density regions on scales below the free-streaming length scale, and therefore suppress the perturbations in the
small-scale matter spectrum. Both the free-streaming of neutrinos and
the suppression of small-scale power due to dark energy are late-time
effects. However, their effects may be disentangled in either the
CMB lensing power spectrum~\cite{Kaplinghat:2003bh,Lesgourgues:2005yv}, shear
power spectrum~\cite{Cooray:1999rv,Abazajian:2002ck,Hannestad}, or galaxy power 
spectrum~\cite{Hu:1997mj, Eisenstein:1998hr, Hannestad:2002cn, Lesgourgues:2004ps} 
(moreover see Ref.~\cite{Namikawa:2010re} for a combined lensing analysis). 
We revisit this issue including the presence of early dark energy
because we expect significant degeneracy between the effects of
massive neutrinos and early dark energy~\cite{dePutter:2009kn}.

Previous efforts to constrain models of early dark energy have
focused on measurements of the CMB temperature and
polarization fields, large and small scale structure formation,
Lyman-$\alpha$ forest, Gamma-Ray Bursts, and SNe~\cite{DorRobWet,
  Caldwell, XiaViel, GroSpr, FraLewLin1, FraLewLin2}. In recent work,
the influence of weak gravitational lensing observations of galactic
sources at low redshifts and the CMB as a large redshift source has
also been investigated~\cite{dePutter:2009kn, Hollenstein:2009ph}. In
this work, we explore how well a generic model for dark energy that
does not vanish at early times can be constrained in conjunction with
massive neutrinos from a comprehensive array of next-generation weak
lensing, galaxy, CMB (including lensing), and SN observations,
including all cross-correlations between the non-SN probes.  We
include SNe to help break the degeneracies with the 
late-time dark energy parameters. 
It is worth noting that the comprehensive inclusion of
cross-correlations is new even for the case with just late-time dark
energy. 

We take our fiducial cosmological model to be flat with a nonzero
fraction of dark energy at high redshift, $\Omega_e = 0.01$, as
discussed in the next section. We assume that the dynamics is
described by a canonical scalar field. In agreement with WMAP
data~\cite{Dunkley, Komatsu}, we set  $\Omega_{d0} = 0.742$, $w_0 =
-1.0$, (present dark energy density and equation of state),
${\Omega_c}h^2 = 0.11$ (density of cold dark matter), ${\Omega_b}h^2 =
0.0227$ (density of baryons), $\Omega_{\nu}h^2 = 1.81 \times 10^{-3}$
(density of neutrino), ${\Omega_k} = 0$ (flatness), $n_s = 0.963$
(tilt of power spectrum), ${\rm d}{n_s}/{\rm   d}\ln{k}= 0$ (running
of tilt),  $\Delta_R^2 = 2.21 \times 10^{-9}$ (normalization of power
spectrum), $\tau = 0.087$ (optical depth to Thomson scattering),
$N_{\rm eff} = 3.04$ (effective number of neutrinos). 

The fiducial value of $\Omega_{\nu}h^2$ corresponds to
$\sum{m_{\nu}} = 94~\Omega_{\nu} h^2 {\rm eV} = 0.17~$eV. For the inverted
hierarchy, this value for the sum of neutrino masses actually implies 2
mass eigenstates with masses about 0.065 eV each and a lighter
eigenstate. We have approximated this as 2 massive neutrino
eigenstates (with masses 0.085 eV each) and one massless neutrino
eigenstate. We included the fraction of $\sum{m_{\nu}}$
in one of the two massive eigenstates as a parameter $f_{\nu1}$ but found that
the Fisher matrix constraint on that fraction to be $\sim1/2$ with all 
the cosmological probes and parameters included. Therefore, our
results are not expected to change if, for example, we split
$\sum{m_\nu}$ among 3 mass eigenstates. For brevity, we do not show
this fraction parameter in the tables and figures. Even though
cosmology is primarily sensitive to the sum of neutrino mass
eigenstates, it can probe the neutrino mass  hierarchy in the sense
that if the sum of the neutrino mass is below 0.095 eV (see
Fig.~\ref{fig:numixing}), then the normal hierarchy is the only
solution~\cite{Lesgourgues:2006nd}.    

It is important to note that our cosmological model above includes
information about reionization only through the optical depth. This is
sufficient if reionization happens rapidly enough compared to the age
of the universe during reionization. However, the reionization process
could be more gradual or more complicated (for example, occur in two
stages) and this would manifest itself in a range of effects on the
large-angle CMB polarization even when the optical depth is 
fixed~\cite{Kaplinghat:2002vt}. As a result, the estimate of how well
$\tau$ can be measured is dependent on the features of 
global reionization~\cite{Kaplinghat:2002vt,Holder:2003eb,Hu:2003gh}. 

The constraint on $\tau$ determines how well the amplitude of the primordial
power spectrum is measured such that
$\sigma(\Delta_R^2)/\Delta_R^2=2\sigma(\tau)$, where $\sigma(X)$
denotes the uncertainty in parameter $X$. For constraints on
parameters such as neutrino mass, the knowledge of
the overall power spectrum normalization is
important~\cite{Kaplinghat:2003bh}. Here, we test the effect that 
including a detailed reionization model would have by imposing a floor
on $\tau$ as advocated in 
Refs.~\cite{Holder:2003eb,Hu:2003gh,Kaplinghat:2003bh}. The main
results we quote in this paper, however, assume a sharp reionization
scenario because our main  aim is to see how adding different probes
helps  break degeneracies in the presence of an unknown dark energy
component at high redshift.  

We describe our method to include early dark energy in
Section~2, reviewing the nonlinear matter power
spectrum, weak lensing power spectrum, galaxy power spectrum, SN
distances, and the CMB. In Section~3, we provide prospective early
dark energy and sum of neutrino mass constraints (and potential biases)
obtained from a joint analysis of these probes, based on a Fisher
matrix prescription. Section~4 concludes with a discussion of our
findings. 

\section{Methodology}

We begin with an overview of our calculation. We briefly describe the
EDE cosmological model, and then discuss the relevant observational
variables.  

\subsection{Early Dark Energy}

\begin{figure}[!t]
\epsfxsize=3.4in
\centerline{\epsfbox{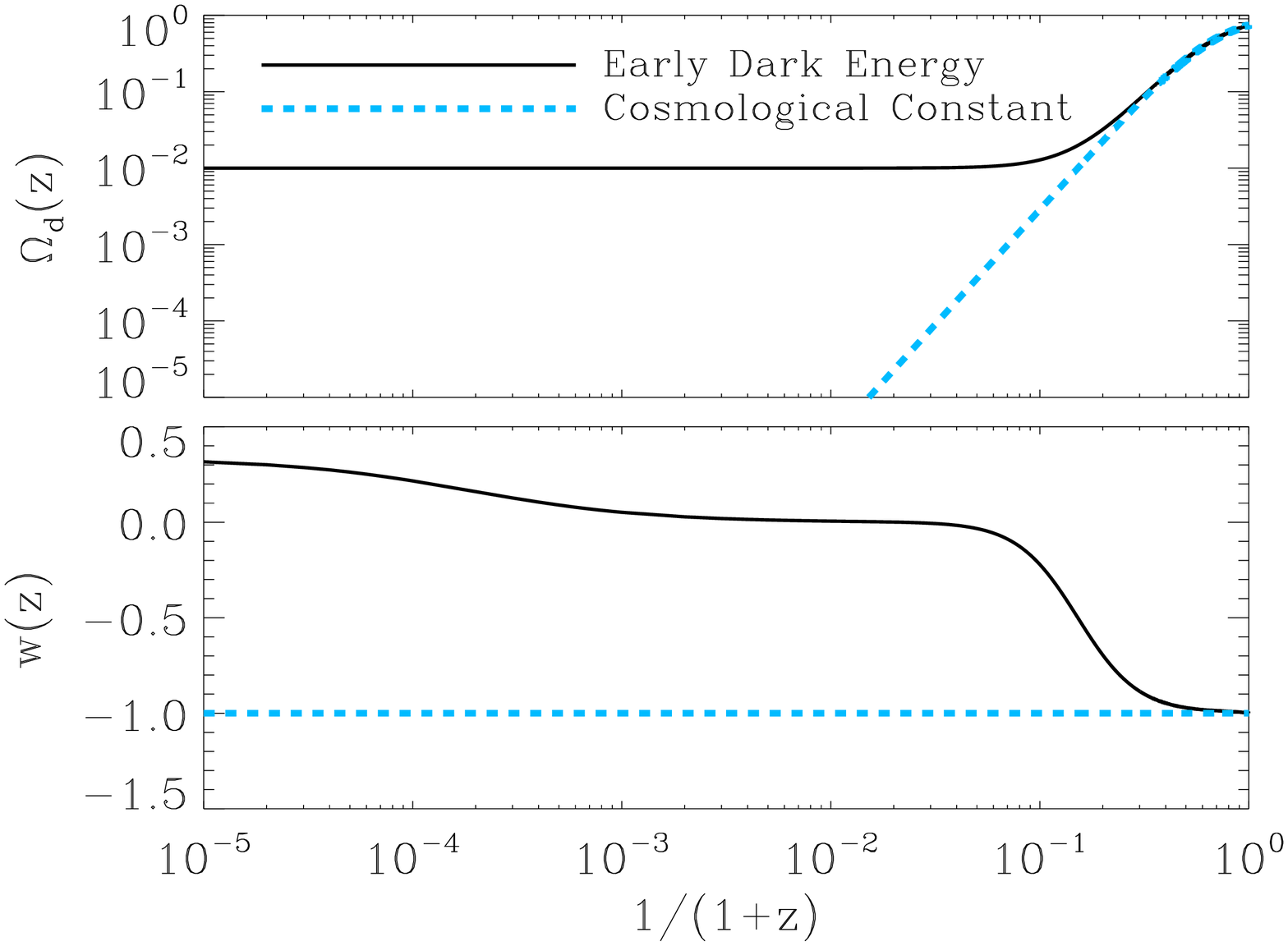}}
\vspace{-1.0em}
\caption{Energy density (top) and equation of state (bottom) of early dark energy 
and a cosmological constant. At low redshifts the EDE mimics a dark energy component 
with the same density and EOS at present, and decouples after redshifts of a few, the exact 
redshift depending on the size of the EDE fraction $\Omega_e$.}
\label{fig:ede}
\end{figure}

Early dark energy changes the expansion rate  and hence cosmological
distances. It also changes the growth of density fluctuations in the
universe and hence the matter power spectrum \cite{Dodelson:2001fq}.

We begin by expressing the expansion rate of the universe in terms of
the dark energy density $\Omega_d(z)$ (in units of the critical
density) as
\begin{equation}
H(z) = H_0 \sqrt{{\Omega_r (1+z)^4 + \Omega_m (1+z)^3 + \Omega_k (1+z)^2} \over {1- \Omega_d(z)}} ,
\end{equation}
where $H_0 = 100 ~ h$ km s$^{-1}$ Mpc$^{-1}$ is the Hubble constant, and 
$\{ \Omega_r, \Omega_m, \Omega_k \}$ are the present radiation, matter, and 
curvature densities in units of the critical density. The
present matter density is further composed of the densities of the
cold dark matter, baryons, and massive neutrinos ($\Omega_m = \Omega_c
+ \Omega_b + \Omega_\nu$). The evolution of dark energy is
conventionally expressed as a function of its equation of state (EOS),
$w(z)$,  
\begin{equation}
\Omega_d(z) = {{\Omega_{d0} H_0^2} \over {H^2(z)}} \exp\left(3 \int_0^z{dz
  {1+w(z) \over 1+z}}\right),
\label{eq:equation3}
\end{equation}
where $\Omega_{d0}=\Omega_d(0)$ is the present density of the dark energy. 

A uniform and constant vacuum density ($w=-1$) is simple but suffers
from the well-known coincidence problem. The value of the dark energy
density has to be fine-tuned so that it only affects the dynamics of
the universe at present. This coincidence 
problem motivates the exploration of solutions other than $\Lambda$CDM 
(e.g.~\cite{Zlatev,Albrecht:1999rm}). Among the possibilities that
allow for $w>-1$ are models in which the evolution of the dark energy
density is such that it is large enough to affect the universal
dynamics even at $z>2$. They may even alleviate the coincidence
problem~\cite{Dodelson:2001fq,Griest:2002cu}.

A realization of early dark energy is given by the "tracker"
parameterization of Doran $\&$ Robbers (2006)~\cite{DorRob}, 
where the dark energy tracks the dominant component in the
universe. For this case, it is simpler to parameterize the dark energy
density evolution directly, rather than express it in terms of an
evolving equation of state. We use a modified form of the Doran and
Robbers (2006) \cite{DorRob} parameterization that tracks the equation
of state of the dominant energy, as shown in Fig.~\ref{fig:ede},   
\begin{eqnarray}
\Omega_d(z) &=& \Omega_{d0}{(1+z)^{3+3w_0} \over h_w^2(z)} \nonumber\\
&+& \Omega_e v(z)\left(1-{(1+z)^{3+3w_0} \over
    h_w^2(z)}\right), \label{eq:equation2} \\
h_w^2(z)  & = & \Omega_{d0}(1+z)^{3+3w_0}+ \Omega_m (1+z)^3
\nonumber\\
&+& \Omega_r(1+z)^4 + \Omega_k (1+z)^2 ,\nonumber
\end{eqnarray}
where $w_0 = w(0)$. The function $v(z)$ should have the properties that it
asymptotes to unity at large redshift and $v(0)=0$, thus ensuring that
$\Omega_d(z)$ asymptotes to $\Omega_e$ at large redshift and
$\Omega_d(0)=\Omega_{d0}$. We use $v(z)=1-(1+z)^{3w_0}$
\cite{DorRob}, but any other parameterization such that
$d\ln(v)/d\ln(z)={\cal O}(1)$ will give similar results. 
Note that the first term proportional to $\Omega_{d0}$ is dark energy
density as a function of redshift for a model with present density of
dark energy $\Omega_{d0}$ and constant EOS $w_0$. Thus, in this 
parameterization with early dark energy, the effect at low redshift is
the same as a model with constant EOS model $w_0$ and density
$\Omega_{d0}$.
Quantitatively, the $\Omega_e$ term (``early dark energy'') in
Eqn.~\ref{eq:equation2} constitutes $[0, 2.1, 8.0, 17.7]$\%  at
redshifts $z = [0, 1, 2, 3]$ respectively of the overall amount of
dark energy $\Omega_d(z)$ for $w = -1$ and $\Omega_e = 0.01$.  

We may compute the EOS using the expression $w(z) = {-1 + {{(1+z)}
    \over z} {{{d \ln [\Omega_d(z) H^2(z)]} \over {3~d \ln
        z}}}}$. At z=0, $w(z)=w_0$ and increases with z, tending to 0
if the dominant component of energy density is due to pressureless
matter and an example of this behavior is shown in Fig.~\ref{fig:ede}. 

\subsection{Matter Power Spectrum}

\begin{figure}[!t]
\epsfxsize=3.3in
\centerline{\epsfbox{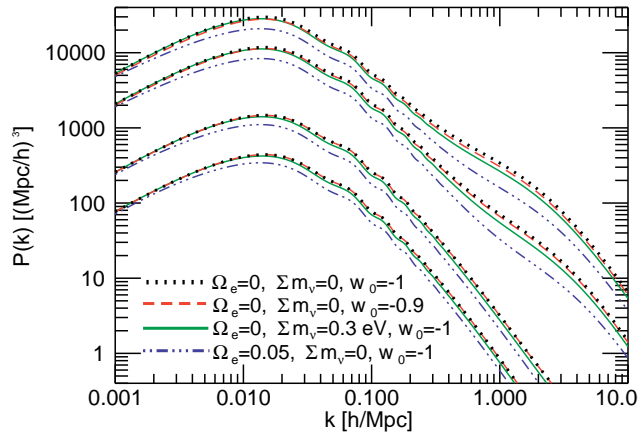}}
\vspace{-1.0em}
\caption{Matter power spectrum $P(k)~(({\rm Mpc}/h)^3)$ against wavenumber $k~(h/{\rm Mpc})$ at 
$z = [0,1,5,10]$ (high to low) in four distinct cosmologies: $\Lambda$CDM (dotted black), 
wCDM (dashed red, $w_0 = -0.9$ and $\Omega_e=0$), $\Lambda$CDM with massive neutrinos (solid
  green, $\sum m_\nu = 0.3$ eV), and a CDM universe with EDE
  (dot-dashed blue, $\Omega_e = 0.05$, $w_0=-1$).}
\label{fig:pk}
\end{figure}

Given a particular parameterization of the EOS, the growth history
depends on the underlying microphysical model for dark energy. In this
paper, we will assume that the underlying model is a single scalar
field. Then each $\Omega_d(z)$ or equivalently $w(z)$ can be connected
to an underlying scalar field potential and equations for
perturbations in matter, radiation and dark energy may be 
written down (e.g.~\cite{Caldwell:1997ii,Ma:1999dwa, Skordis:2000dz}.)
This allows the calculation of the growth of matter perturbations. 
In order to correctly calculate the linear matter power spectrum, we
numerically solve the Boltzmann equations with a modified 
version of CAMB~\cite{LCL}.  We approximate the effect of an EDE
component on the linear matter power spectrum by use of the PPF module
by W.~Fang \cite{camb, Fang}.

We extend this power spectrum to nonlinear scales by calculating the
appropriate effective spectral index, effective spectral curvature,
and nonlinear scale, and employing the fitting functions provided in
Smith et al.~(2001)~\cite{Smith}.  
The underlying cosmology in the Smith et al. fitting function manifests itself
in two distinct ways. First, cosmology impacts the evolution of the matter
density, $\Omega_m(z)$, and the evolution of the growth of matter perturbations,
$D(z)$. The cosmology then fixes the functional form and coefficients associated
with the fitting functions, which are fine-tuned to a suite of $\Lambda$CDM
N-body simulations. 

Thus, whereas an arbitrary dark energy EOS could
leave its imprint on the nonlinear matter power spectrum via its
influence on $\Omega_m(z)$ and $D(z)$, the cosmological dependence of
the N-body fitting functions remain fine-tuned to a $w\equiv-1$ dark
energy EOS. For this reason, the nonlinear solution for a
non-$\Lambda$CDM cosmology is only approximate, and studies have shown
that it could lead to an underestimation of dark energy constraints:
for example, in prospective weak lensing measurements by factor of
two~\cite{McDonald, Joudaki}. Our use of this nonlinear extension 
could also underestimate the suppression of the matter power spectrum
on small scales due to massive neutrinos (e.g. Ref~\cite{Saito:2008bp}). 
Our constraints on neutrino mass, in particular from weak lensing, may therefore 
be conservative. Further work is needed to quantify this effect to the level of 
precision required for the next-generation telescopes considered in
this paper (however, also see Refs.~\cite{Brandbyge:2008rv,Wong:2008ws,Saito:2009ah}).

Fig.~\ref{fig:pk} illustrates the matter power spectrum in four
distinct cosmologies: $\Lambda$CDM, wCDM, $\Lambda$CDM with massive
neutrinos, and a CDM universe with EDE. The inclusion of massive
neutrinos induces a suppression of the matter power spectrum on scales
below the free streaming length and this is, to an extent, degenerate
with the suppression introduced by early dark energy. The suppression
due to the presence of early dark energy is evident in
Fig.~\ref{fig:pk} and is a result of the fact that the universe is
expanding faster compared to a model with no early dark energy. 
The present ($\Omega_{d0}$) and early ($\Omega_e
\neq 0$) dark energy densities are further degenerate with the present
DE equation of state $w_0$.
A combination of large and small scale probes are therefore needed to
break the degeneracy between these cosmological parameters $\(m_\nu,
\Omega_{d0}, w_0, \Omega_e\)$~\cite{Dodelson,Hannestad,Caldwell,Hollenstein:2009ph}.  

\subsection{Weak Lensing Tomography}
\label{subforma}

The images of distant galaxies are gravitationally lensed by matter
inhomogeneities along the line-of-sight. In the weak lensing regime these
percent-level magnifications and shape distortions of galaxies need to be
analyzed statistically (see~\cite{Refregier,Schneider} for a
review). By extracting the shear power spectrum of weakly lensed  
sources \cite{Wittman, Jarvis, Hoekstra, KaiserWL, BaconRE, Waerbeke,
  Brownetal, WaerbekeMH, Heymansetal}, the nature of the dark energy
has been constrained with lensing surveys~\cite{Jarvis, Hoekstra}. An
important aspect is that the lensing power spectrum depends on both
the lensing kernel and the growth of perturbations, making lensing the
most powerful probe of the underlying cosmology~\cite{Albrecht}. 

\begin{figure}[!t]
\vspace{-0.8em}
\epsfxsize=3.4in
\centerline{\epsfbox{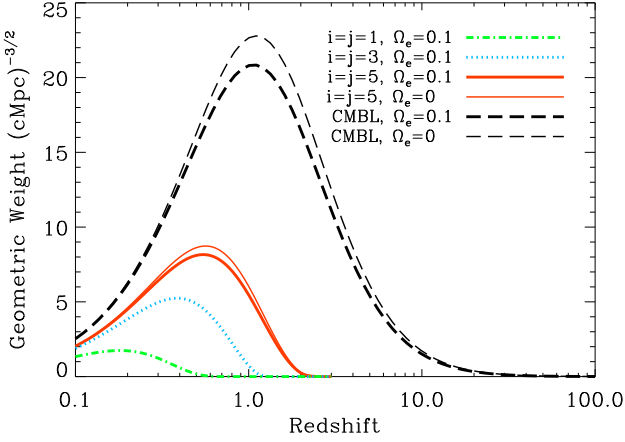}}
\vspace{-0.8em}
\caption{Redshift dependent geometric weight
$W_{ij}(z) = {{9 \over 16}{(\Omega_m H_0^2)^2}H(z)\chi^{7/2}(z){\zeta}^{\kappa}_{i}(z){\zeta}^{\kappa}_{j}(z)}$
  (see text following Eqn.~\ref{eqkappa}) of CMB weak lensing (CMBL) and tomographic low redshift weak lensing bins
for a flat CDM model with $\Omega_e = 0.1$ and $w_0=-1.0$. For CMB lensing and the fifth tomographic bin we 
also plot the kernel for $\Omega_e=0$. The lensing kernel
captures approximately the redshift dependence of the
integrand for the lensing power spectrum (including that from the
matter power spectrum). The diminishment of the kernels for
$\Omega_e>0$ stems from the  increase of $H(z)$ with increasing
$\Omega_e$.}  
\label{fig:weights}
\end{figure}

We employ weak lensing tomography, as this provides information about
the redshift distribution of the intervening lenses, and thereby
allows for more stringent constraints on cosmological
parameters~\cite{Hu99,MaHu}.   
We work within the Born
approximation~\cite{CooHu,HirSel,ShaCoo} to 
compute the lensing potential (weighted projection of the
three-dimensional gravitational
potential~\cite{Kaiser:1991qi,Hu:2000ee}):  
\begin{equation}
\phi_i({\bf \hat{n}}) = \int_0^{z_{\rm max}} dz
K(z) {{{\zeta}_i^{\kappa}(z)}} \Phi({\bf \hat{n}},z), 
\label{eq:zz9}
\end{equation}
where $K(z) \equiv H(z) d\chi/dz = 1$ in a flat universe. The
lensing weight function (see Fig.~\ref{fig:weights}) of the $i^{\rm th}$
tomographic bin is given by 
\begin{eqnarray}
& & {\zeta}_i^{\kappa}(z) = {-2 H^{-1}(z)\over {\bar{n}_i\chi(z)}}
\int_{\max(z,z_i)}^{z_{i+1}} dz_s {{X(z,z_s) \rho(z_s)} \over {\chi(z_s)}}, \label{eq:zz9b} \\
& & X(z_o,z_s) = {H_0^{-1} \over {\sqrt{| \Omega_k |}
  }} S_k\left({\sqrt{| \Omega_k |}} \int_{z_o}^{z_s} dz'{H_0 \over
    H(z')}\right), 
\end{eqnarray}
where 
$X(z_o,z_s)$ is the comoving distance to
object at redshift $z_s$ measured by observer at redshift $z_o$, such that $\chi(z) \equiv X(0,z)$, and
$S_k(x) = \[\sin(x), \sinh(x), x\]$ for a $\[{\rm closed, open, flat}\]$ universe, respectively. 
We set $\zeta_i^\kappa(z)=0$ explicitly if  $z>z_{i+1}$. The quantity
$\rho(z_s)$ contains   the source galaxy distribution, the integral
over which is $\bar{n}_i$ (see Sec.~\ref{subsa}).   
For the lensing of the CMB photons, we replace the galaxy
source distribution with a $\delta$-function at the last scattering
surface.  

The cross correlation of the respective Fourier coefficients at
angular multipoles ${\bf l}$ and ${\bf l}^{\prime}$ is given by 
\begin{equation}
{\left\langle{\phi_i^{*}({\bf l})\phi_j({\bf l}^{\prime})}\right\rangle} = (2\pi)^2\delta_{\rm D}({\bf l} - {\bf l}^{\prime}) C_{ij}^{\phi\phi}(\ell) 
\label{eq:ccextra2}
\end{equation}
in the flat sky limit, which through the Limber approximation reduces to~\cite{Limber,Kaiser:1991qi,HuJain}:  
\begin{equation}
C_{ij}^{\phi\phi}(\ell) = \int_0^{z_{\rm max}} dz K(z){{\zeta}_i^{\kappa}(z){\zeta}_j^{\kappa}(z)} {{H(z)} \over {\chi^2(z)}} P_{\Phi\Phi}\left({k,z}\right).
\label{phiphi}
\end{equation} 
This projected spectrum remains the same in the all sky formulation~\cite{Hu:2000ee}.

To find the relation between the power spectrum of the potential
($P_{\Phi\Phi}({\bf k},z)$) and matter perturbations
($P_{\delta\delta}({\bf k},z)$) on sub-horizon scales, we use the
Poisson equation in Fourier space,
\begin{equation}
{\Phi}({\bf k},z) = -{3 \over 2}{\Omega_m}{\left({H_0 \over
      k}\right)}^2\delta({\bf k},z)(1+z). 
\label{eq:zz3}
\end{equation}

Accounting for the definition of two-point correlations of the
potential and density fields:  
\begin{eqnarray}
\left\langle {\Phi^{*}({\bf k},z)\Phi({\bf k}^{\prime},z)} \right\rangle = (2\pi)^{3} \delta_{\rm D}({\bf k} - {\bf k}^{\prime}) P_{\Phi\Phi}({k},z), \nonumber\\
\left\langle {\delta^{*}({\bf k},z)\delta({\bf k}^{\prime},z)} \right\rangle = (2\pi)^{3} \delta_{\rm D}({\bf k} - {\bf k}^{\prime}) P_{\delta\delta}({k},z) , 
\label{eq:zz4}
\end{eqnarray}
we obtain the power spectrum of the potential in terms of the power
spectrum of density fluctuations as
\begin{equation}
P_{\Phi\Phi}({k},z) = {9 \over 4} \Omega_m^2 (1+z)^2
{\left({H_0 \over k}\right)}^4 P_{\delta\delta}({k},z). 
\label{eq:cc6pp}
\end{equation}
Reexpressing the power spectrum of the matter as a dimensionless quantity,
\begin{equation}
{\Delta_{{\delta\delta}}^2(k,z)} = k^3 P_{{\delta\delta}}(k,z) /(2\pi^2) ,
\label{eq:eqnpk}
\end{equation}
the spectrum of the convergence field, $C_{\ell}^{\kappa\kappa} = (1/4)\ell^2(\ell+1)^2 C_{\ell}^{\phi\phi}$, is given by:
\begin{eqnarray}
\label{eqkappa}
& & C^{\kappa\kappa}_{ij}(\ell) = A_{\ell}^{\kappa\kappa}
  \int_0^{z_{\rm max}} dz K(z)
  {\zeta}^{\kappa}_{i}(z){\zeta}^{\kappa}_{j}(z) ~ \times \\ 
&&\hspace*{1cm} {(1+z)^2H(z)\chi^5(z)}\Delta_{{\delta\delta}}^2\left({\ell+1/2 \over \chi(z)},z\right),
\nonumber  
\end{eqnarray}
where the density and scale dependent amplitude 
$A_{\ell}^{\kappa\kappa} = {(9\pi^2/8){(\Omega_m H_0^2)^2}\ell^2(\ell+1)^2}\left(\ell+{1/2}\right)^{-7}$, and $\Delta_{{\rm \delta\delta}}^2({\ell/\chi(z)},z)$ 
encapsulates the full nonlinear dark matter power spectrum. 
We construct a geometric weight $W_{ij}(z)$ that approximately
captures the contribution to the convergence power spectrum from
different redshifts for a broad range of multipoles by removing
a factor of $(1+z)^2\chi^{3/2}(z)$ from the integrand in
Eqn~\ref{eqkappa}. This results in $W_{ij}(z) = {{9 \over 16}{(\Omega_m
    H_0^2)^2}H(z)\chi^{7/2}(z){\zeta}^{\kappa}_{i}(z){\zeta}^{\kappa}_{j}(z)}$. The
reason this approximation takes into account the redshift contribution
to the convergence  power spectrum is that the matter power spectrum
$\Delta_{{\rm \delta\delta}}^2(k,z)$ scales roughly like $k^{3/2}$ for
scales ${\cal O}(0.1/{\rm Mpc})$. 
We plot this geometric weight $W_{ij}(z)$ in Fig.~\ref{fig:weights}, and
this shows that
most of the contribution to low redshift weak lensing comes from
$z\approx0.5$, and in the case of CMB weak lensing
$z\approx1$. 

\begin{figure*}[!t]
\begin{center}
\vspace{-0.8em}
\includegraphics[scale=0.45]{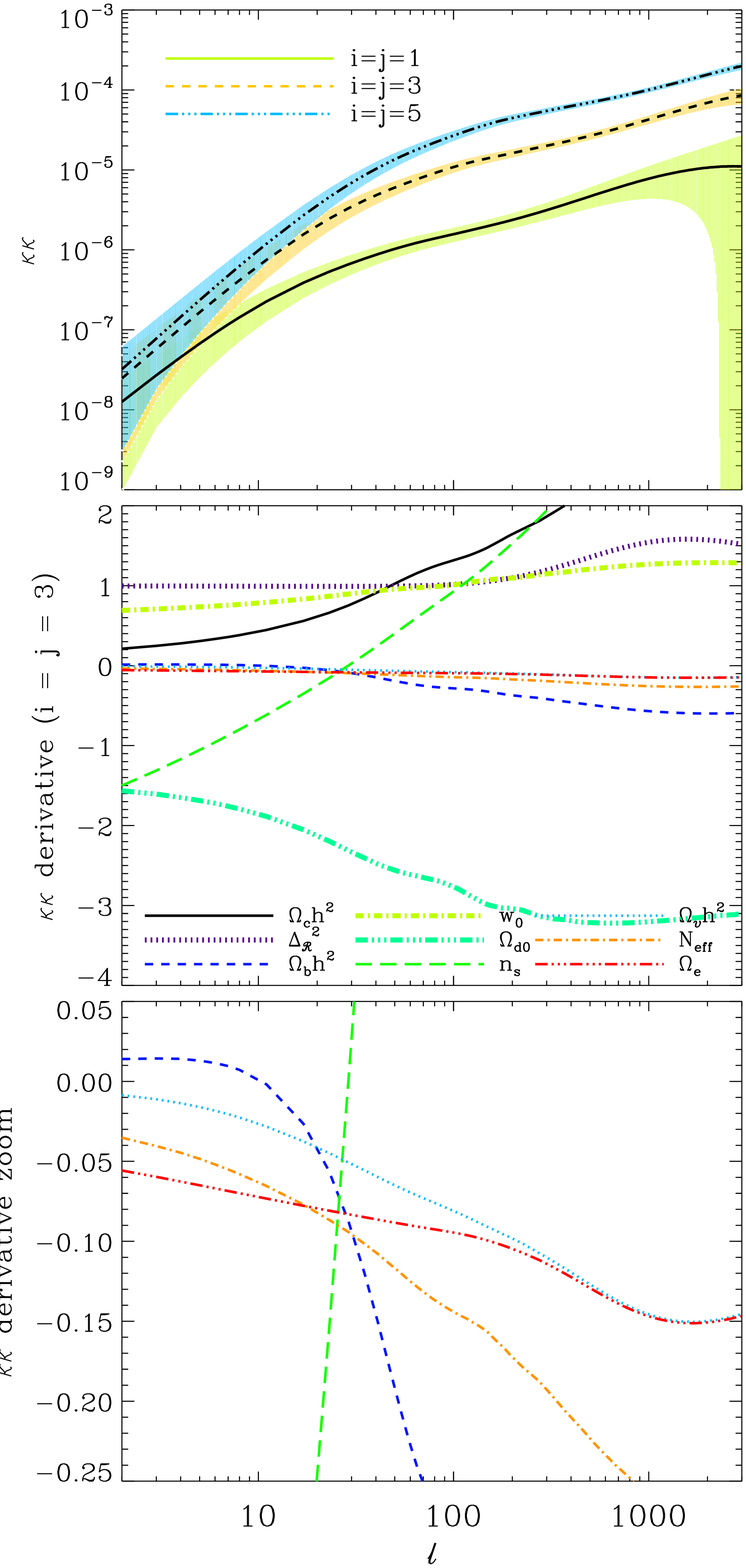}
\hspace{1em}
\includegraphics[scale=0.45]{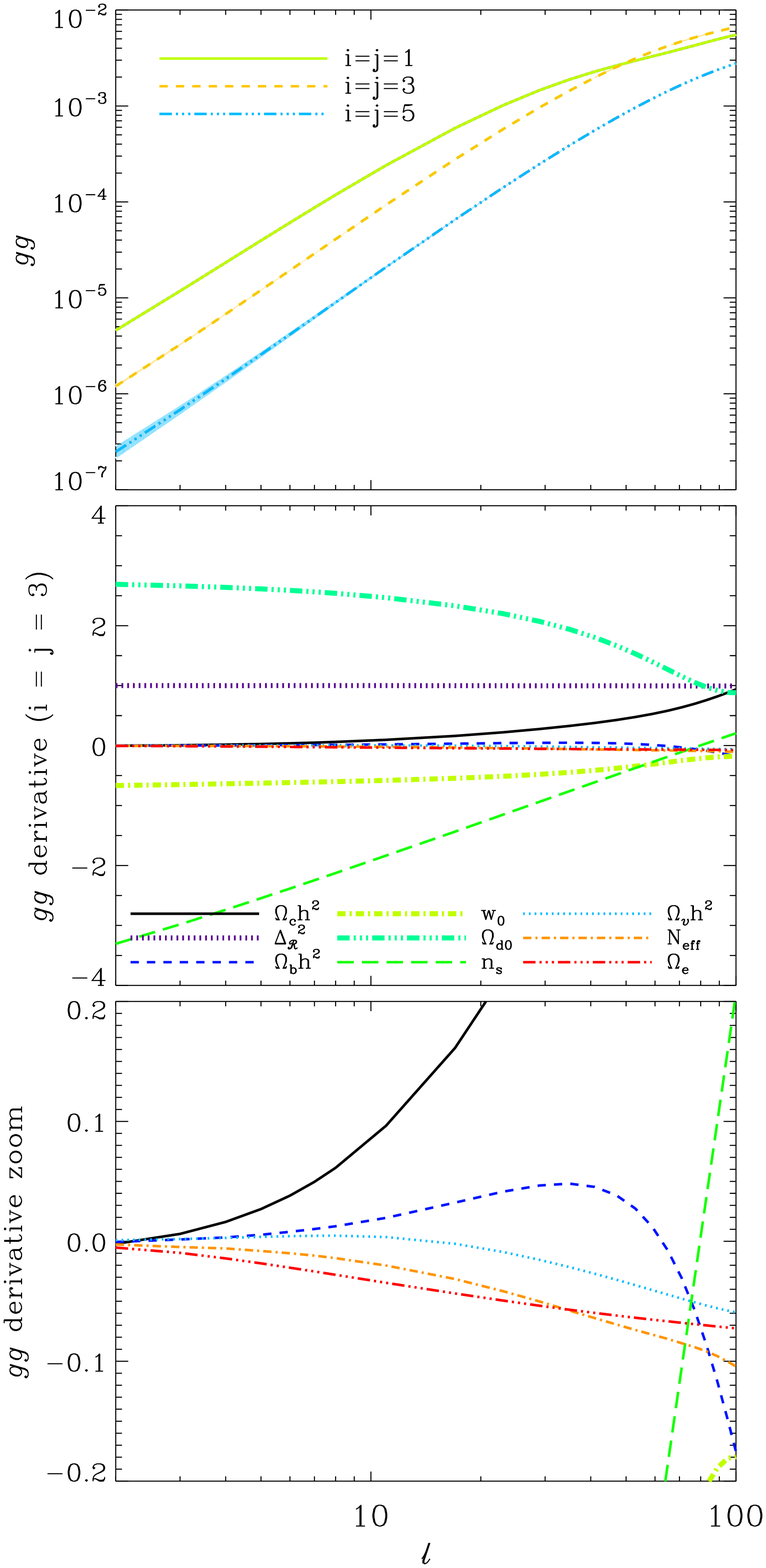}
\end{center}
\vspace{-2em}
\caption{{\it Left}: \underline{Top}: Convergence power spectra
  $\ell(\ell+1)C^{\kappa\kappa}_{ij}/{2\pi}$ for the case of five
  tomographic bins in the fiducial cosmology. We include the expected
  noise for LSST as a band about the curve. \underline{Mid}: Logarithmic
  derivative $d\ln{C^{\kappa\kappa}_{ij}}/d\ln{p_k}$ of the
  convergence power spectrum with cosmological parameters $p_k$ for
  the third tomographic bin ($i = j = 3$). The derivatives of the
  other tomographic bins have similar shapes. \underline{Bottom}: The
  sub-window zooms in on the logarithmic derivatives with 
  sum of neutrino masses (dotted blue) and EDE density (dot-dot-dashed red). 
{\it Right}: Same as {\it Left} but for galaxy power
spectra. The noise contribution (both LSST and JDEM) is at most on sub-percent level and
decreases towards smaller scales, as the noise is constant with scale
whereas $C_\ell^{gg}$ increases with scale. This is to be contrasted
with lensing tomography where $C_\ell^{\kappa\kappa}$ peaks at around
$\ell = 10$ and thereafter decreases continuously.}
\label{fig:wlbins}
\end{figure*}

Fig.~\ref{fig:wlbins} further illustrates the effect of
varying the dark energy and other parameters on the correlation
function in the third tomographic bin. The shapes are roughly the same
across the other bins. The similarity in the shapes of the variation
due to early dark energy and sum of neutrino masses implies that there will be
a degradation in the constraints on the sum of neutrino masses due to the addition
of early dark energy. 

\subsection{Galaxy Angular Power Spectrum}
\label{galpow}

\begin{figure*}[!t]
\begin{center}
\vspace{-0.8em}
\includegraphics[scale=0.45]{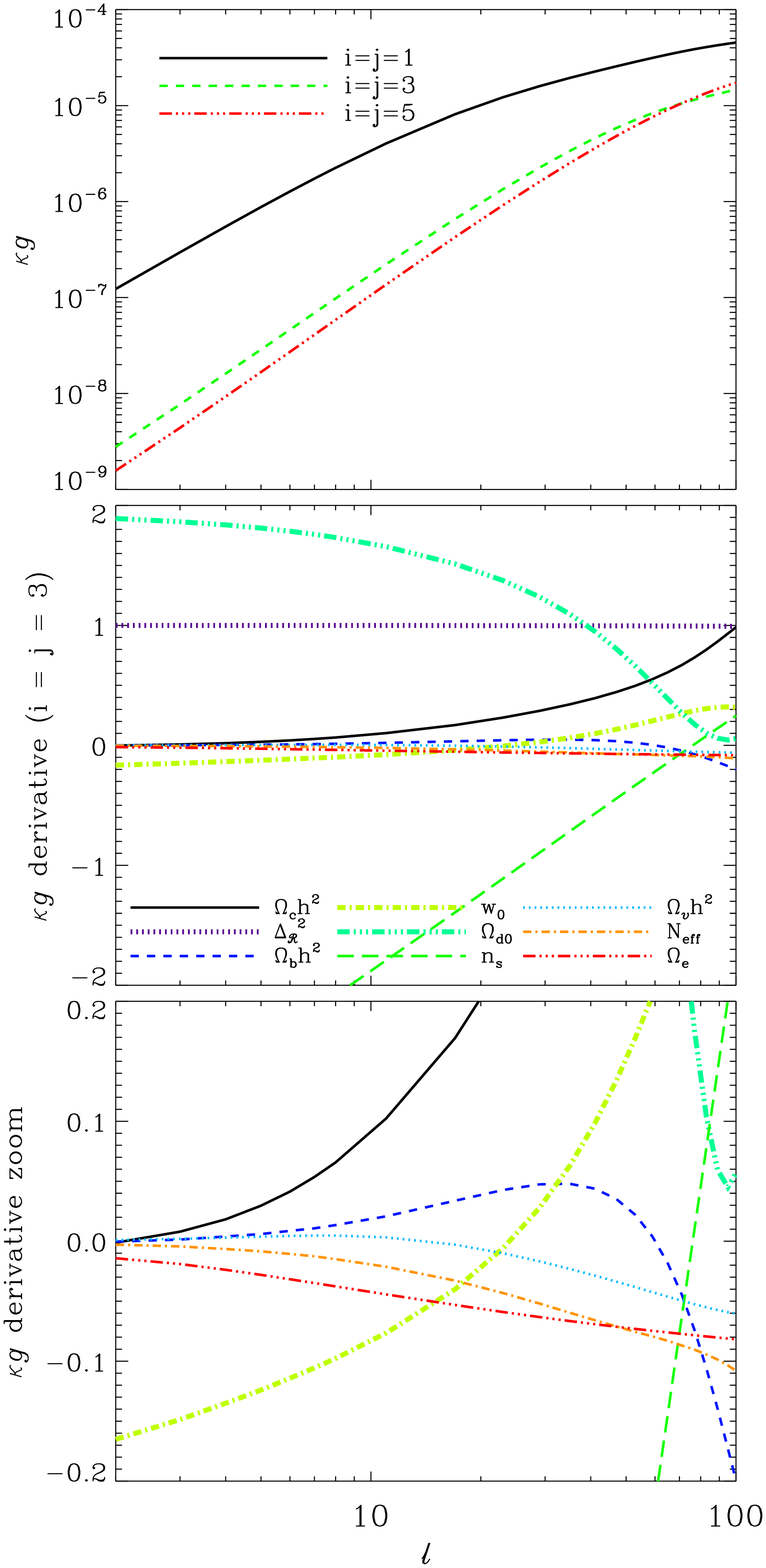}
\includegraphics[scale=0.445]{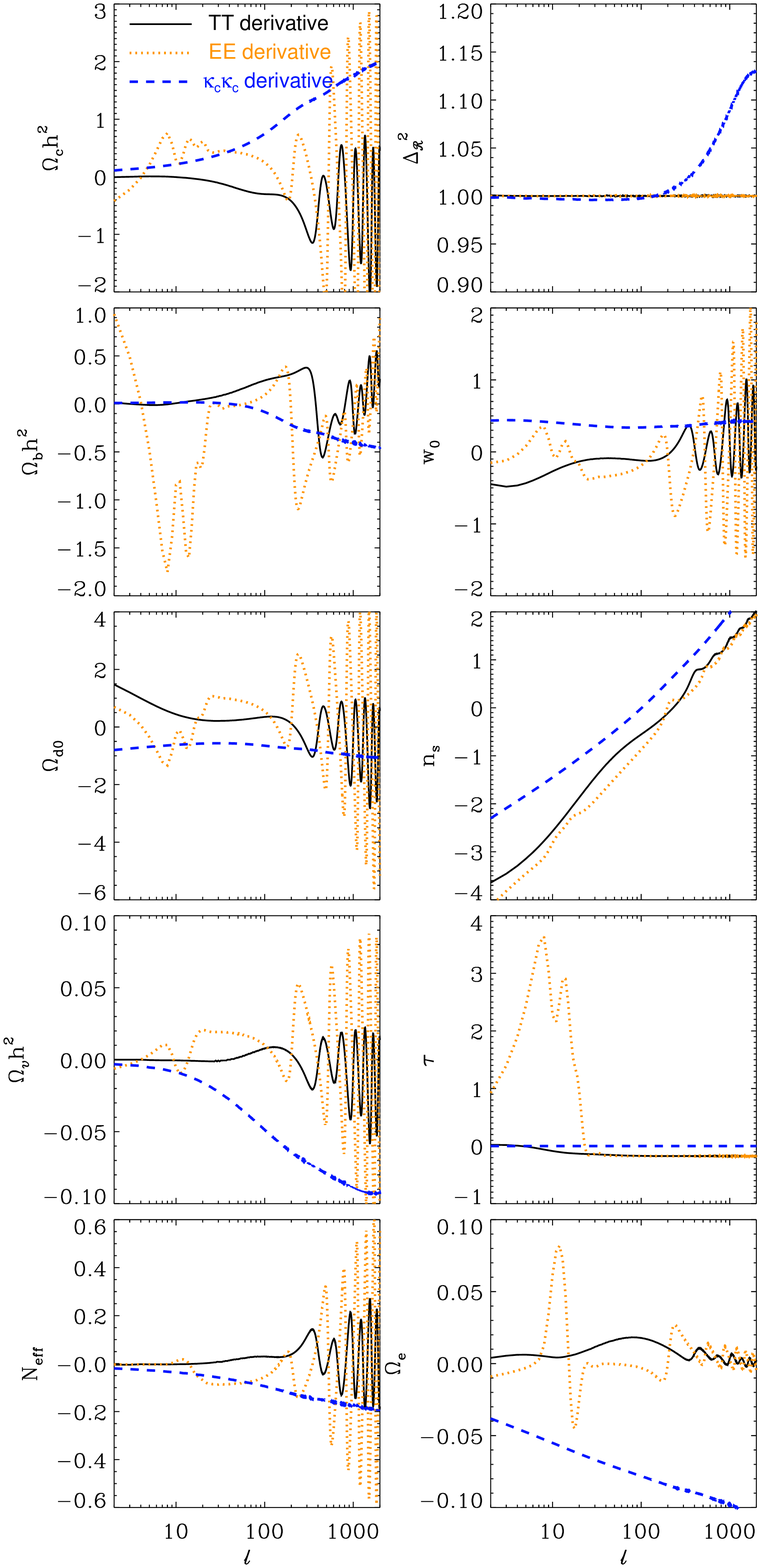}
\end{center}
\vspace{-2em}
\caption{{\it Left}: \underline{Top}: Tomographic galaxy-lensing
  correlations $\ell(\ell+1)C^{{\kappa}g}_{ij}/{2\pi}$. Here we only illustrate a subset of 
  cases where galaxy and lensing bins fully overlap. \underline{Mid}: Logarithmic 
  derivatives of the power spectra $d\ln{C^{{\kappa}g}_{ij}}/d\ln{p_k}$ with 
  parameters $p_k$ for the third tomographic bin in both galaxy and lensing
  (similar characteristics for other bins). \underline{Bottom}: Zooming in on the
  derivatives with EDE density and sum of neutrino masses. {\it Right}:
  Logarithmic derivatives of the CMB temperature (solid black),
  E-mode (dashed blue), and lensing potential (dotted red) power
  spectra with cosmology.} 
\vspace{-1.4em}
\label{fig:galwl}
\end{figure*}

\begin{figure*}[!t]
\begin{center}
\vspace{-0.8em}
\includegraphics[scale=0.44]{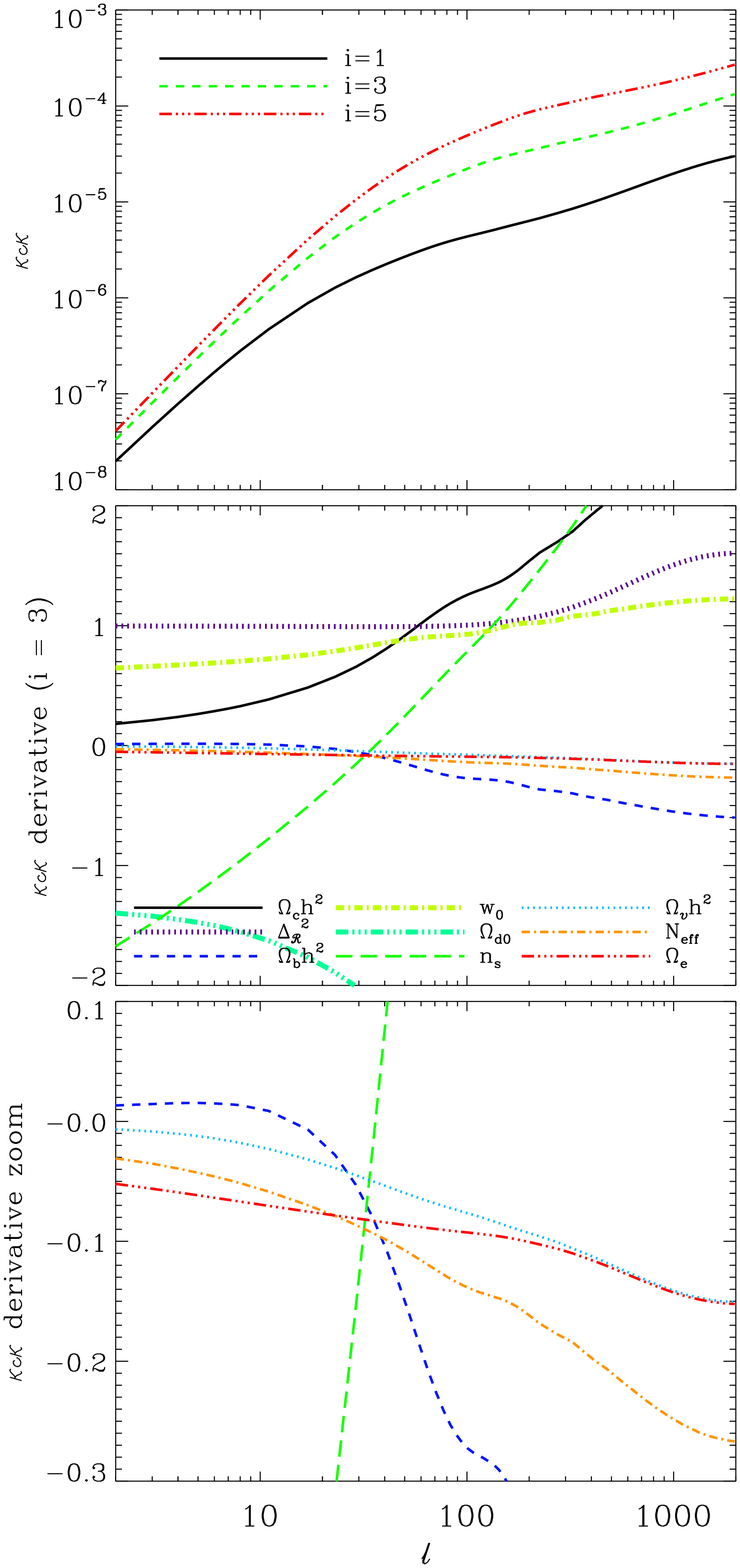}
\hspace{1em}
\includegraphics[scale=0.44]{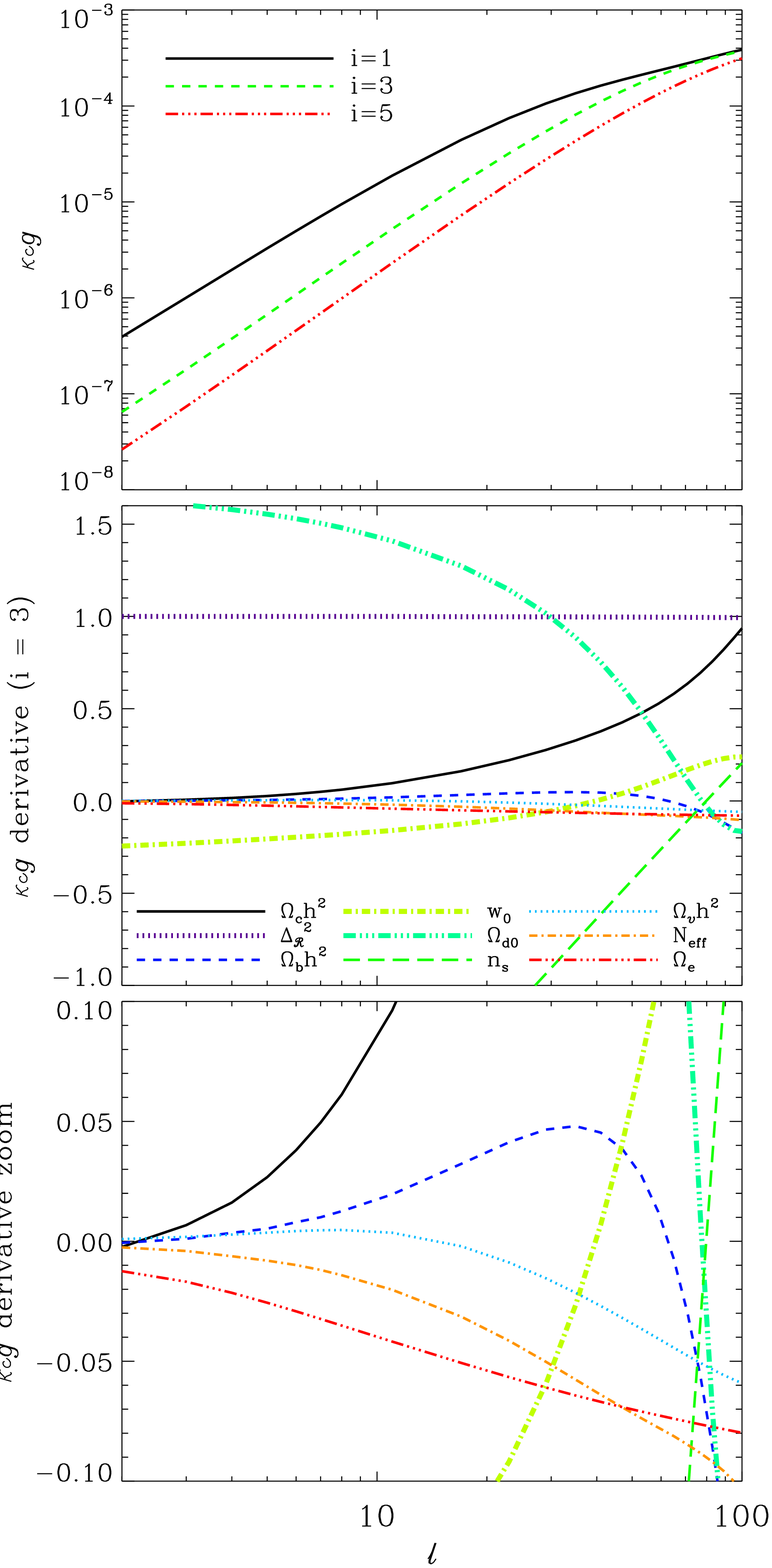}
\end{center}
\vspace{-2.1em}
\caption{{\it Left}: \underline{Top}: Correlations between weak lensing of the CMB and low-redshift sources $\ell(\ell+1)C^{{\kappa_c}\kappa}_{i}/{2\pi}$ for three of the five tomographic bins. \underline{Mid}: Logarithmic derivatives of the power spectrum $d\ln{C^{{\kappa_c}\kappa}_{i}}/d\ln{p_k}$ with parameters $p_k$ for the third tomographic bin (similar characteristics for other bins). \underline{Bottom}: Zoom on derivatives with EDE density and sum of neutrino masses.
{\it Right}: Same as {\it Left} but for correlations between galaxy tomography and lensing of the CMB.}
\vspace{-0.8em}
\label{fig:cmblenscorrs}
\end{figure*}

\begin{figure*}[!t]
\begin{center}
\vspace{-0.4em}
\includegraphics[scale=0.45]{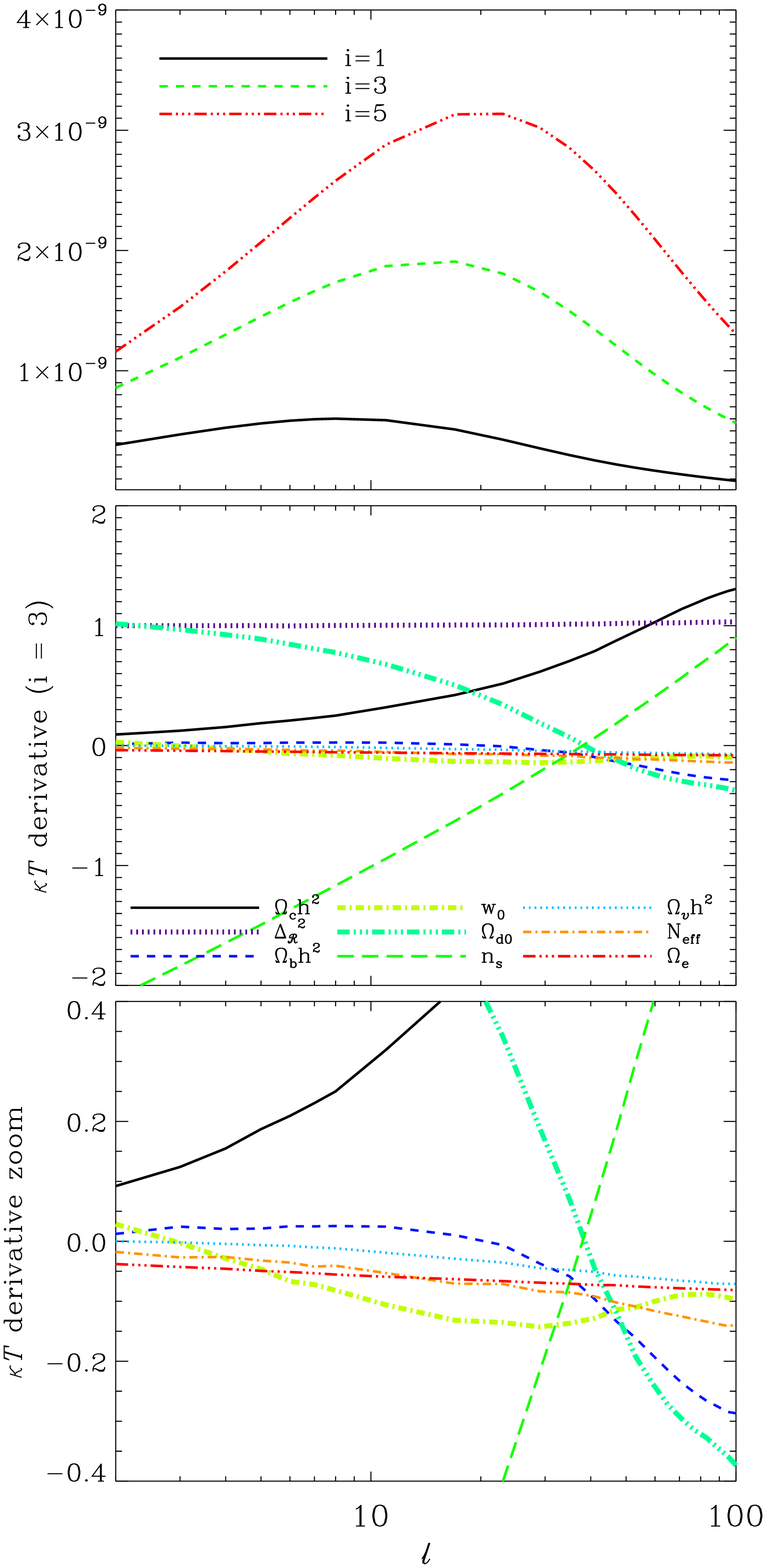}
\hspace{1em}
\includegraphics[scale=0.45]{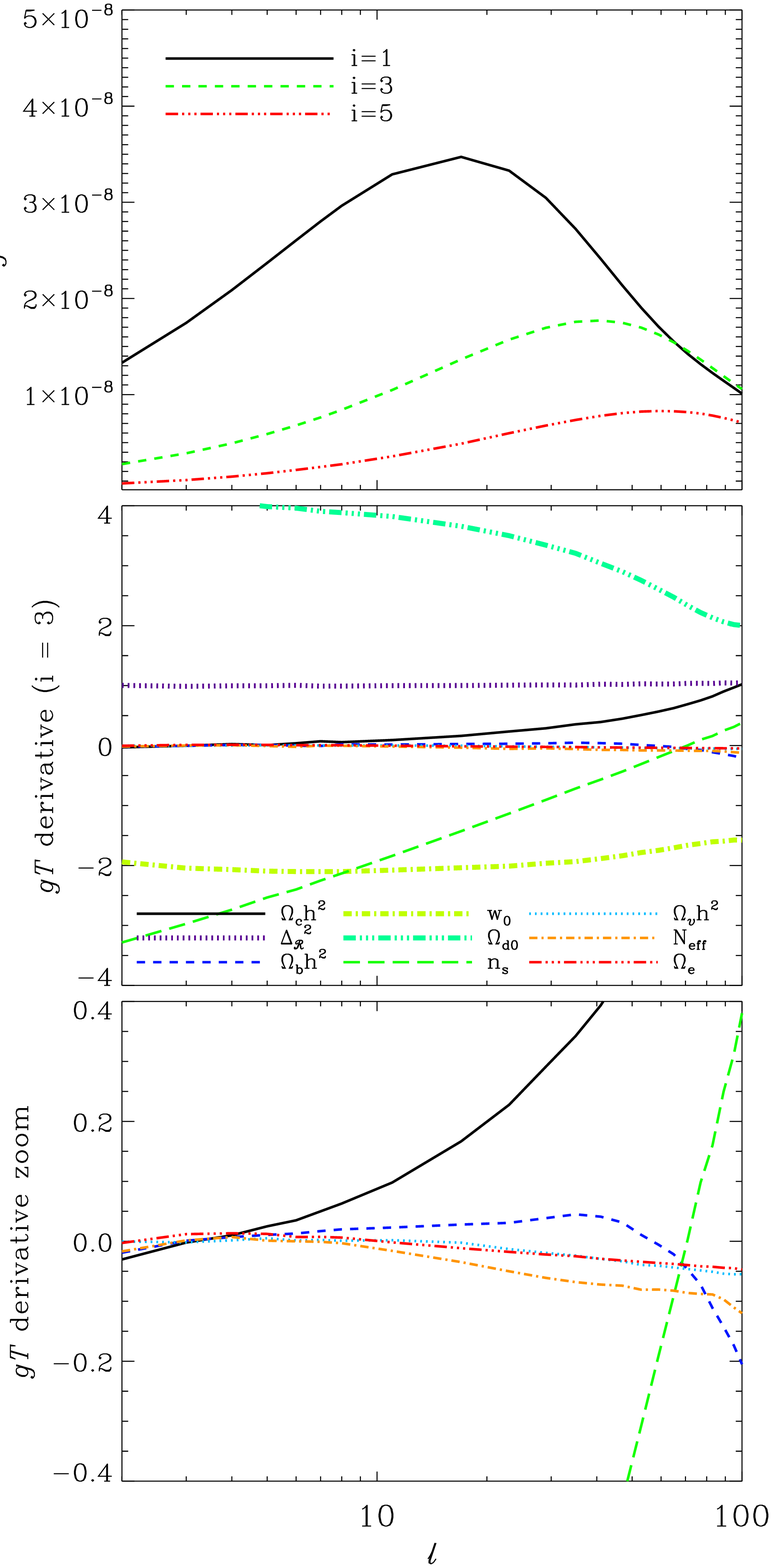}
\end{center}
\vspace{-2em}
\caption{{\it Left}: \underline{Top}: ISW-WL correlations $-\ell(\ell+1)C^{{\kappa}T}_{i}/{2\pi}$ for three of the five tomographic bins. \underline{Mid}: Logarithmic derivatives of the power spectrum $d\ln{C^{{\kappa}T}_{i}}/d\ln{p_k}$ with parameters $p_k$ for the third bin (similar characteristics for other bins). \underline{Bottom}: Zoom on the EDE and neutrino mass derivatives. {\it Right}: Same as {\it Left} but for ISW-galaxy correlations.}
\vspace{-0.5em}
\label{fig:wlisw}
\end{figure*}

The galaxy-galaxy angular power spectrum $C^{gg}(\ell)$ can be
obtained analogously to the derivation of $C^{\phi\phi}(\ell)$ in
Eqn.~\ref{phiphi}. The two-dimensional projection of the
three-dimensional biased tracer of the density field $\delta^g({\bf
  \hat{n}},z)$ is 
\begin{equation}
g_i({\bf \hat{n}}) = \int_0^{z_{\rm max}} dz K(z) {{{\zeta}_i^{g}(z)}} \delta^g({\bf \hat{n}},z).
\label{blablaz}
\end{equation}
From $\left\langle {g^{*}({\bf k},z){g}({\bf k}^{\prime},z)}
\right\rangle = (2\pi)^{3} \delta_{\rm D}({\bf k} - {\bf k}^{\prime})
P_{g{g}}({k},z)$, we thereby obtain the projected galaxy power
spectrum in the Limber approximation:
\begin{equation}
C_{ij}^{gg}(\ell) = \int_0^{z_{\rm max}} dz K(z) {{\zeta}_i^{g}(z){\zeta}_j^{g}(z)} {{H(z)} \over {\chi^2(z)}} P_{gg}({k},z) ,
\label{ggeqn}
\end{equation}
where the weights for $z_i<z<z_{i+1}$ are
\begin{equation}
{\zeta}_i^{g}(z) = \rho(z)/\bar{n}_i , 
\label{ggzeta}
\end{equation}
and ${\zeta}_i^{g}(z) = 0$ if $z\notin (z_i,z_{i+1})$. As a result, only the auto-correlations survive the 
integral. In the linear regime, the three-dimensional galaxy power
spectrum, $P_{gg}({k},z)$, is related to the matter power spectrum via the bias $b({k},z)$:
\begin{eqnarray}
\lefteqn{C_{ij}^{gg}(\ell) = A_{\ell}^{gg}\int_0^{z_{\rm max}} dz K(z)
  {{\zeta}_i^{g}(z){\zeta}_j^{g}(z)}}\\  
&&\hspace*{1.1cm} \times~{H(z)\chi(z)} b^2({k},z)
\Delta^2_{\delta\delta}\left({{\ell+1/2}\over{\chi(z)}},z\right) , \nonumber 
\label{ggeqn2}
\end{eqnarray}
where amplitude $A_{\ell}^{gg} =2\pi^2\left({\ell+1/2}\right)^{-3}$. Restricting
the analysis to largely linear scales (see Table~\ref{table:galcutoffs}), we
include a time-varying linear bias by the first-order expansion $b(z)
= b_0 + b_1 z$. A Gaussian prior of 10\% is applied to these two bias  
parameters~\cite{Schneider:2006br,Zhan:2006gi}, and the fiducial
values are set to $[b_0, b_1] = [1.0, 0.8]$~\cite{Weinberg:2002rm}. 

\subsection{Galaxy-Lensing Correlation}
The cross-correlation of the source galaxy distribution and the weak
lensing convergence is: 
\begin{equation}
C_{ij}^{g{\phi}}(\ell) = \int_0^{z_{\rm max}} dz K(z) {{\zeta}_i^{g}(z){\zeta}_j^{\phi}(z)} {{H(z)} \over {\chi^2(z)}} P_{g\Phi}({k},z) .
\label{gkappaCeqn}
\end{equation}
Correlations between tomographic bins where the galaxy bin lies 
behind the lensing bin are zero (resulting in 15 out of 25 non-zero correlations).
The power spectrum of galaxy and potential perturbations 
may be written as: 
\begin{equation}
P_{g\Phi}({k},z) = {3 \over 2} {\Omega_m \over {a(z)}} {\left({H_0 \over k}\right)}^2 b(k,z) r_c(k,z) P_{\delta\delta}({k},z),
\label{gkappaPeqn}
\end{equation}
where we have introduced the correlation coefficient $r_c$ defined
through the relation $P_{g{\delta}}^2({k},z) = r_c^2(k,z)
P_{gg}({k},z)P_{\delta\delta}({k},z)$.   
Then, using $C_{\ell}^{g{\kappa}} = -\ell(\ell+1)
C_{\ell}^{g{\phi}}/2$, we may write the angular power spectrum for
galaxy and weak lensing correlation as:
\begin{eqnarray}
\label{gkappaeqnyo}
& & C_{ij}^{g{\kappa}}(\ell) = A_{\ell}^{g\kappa}\int_0^{z_{\rm max}} dz K(z) {{\zeta}_i^{g}(z){\zeta}_j^{\kappa}(z)} ~ \times\\ 
& &\quad (1+z) H(z)\chi^3(z) b({k},z)r_c({k},z)\Delta^2_{\delta{\delta}}\left({{\ell+1/2}\over{\chi(z)}},z\right), \nonumber
\end{eqnarray}
where we have defined $A_{\ell}^{g\kappa} = -(3\pi^2/2)(\Omega_m
H_0^2)\ell(\ell+1)(\ell+1/2)^{-5}$. 
Restricting the
analysis to linear scales, we fix $r_c \equiv 1$ in our calculations. 

\subsection{Cosmic Microwave Background ($T$, $E$, $\kappa_c$)}

We consider CMB information from the temperature field ($T$), E-mode
polarization ($E$), and weak lensing of the CMB ($\kappa_c$). The
temperature and E-mode polarization power spectra (and their
cross-correlation) are obtained from a modified version of the
Boltzmann code CAMB \cite{LCL}. 

As for lensing of the CMB photons~\cite{Seljak:1995ve, Zaldarriaga:1998ar, Hu99} (also 
see Refs.~\cite{Smidt:2010by,Das:2011ak} for detections of the lensing power spectrum), we calculate it in
the same manner as in Eqn.~\ref{eqkappa} with the source constrained
to be at the redshift of decoupling ($z_{\rm dec}$), such that  
\begin{equation}
{\zeta}^{\kappa_c}(z) = {-2 \over {H(z)\chi(z)}}{\chi(x(z_{\rm dec})-x(z)) \over \chi(z_{\rm dec})}.
\label{eq:weightkappac}
\end{equation}
Hence, aside from
${\zeta}^{\kappa}_{i}(z){\zeta}^{\kappa}_{j}(z) \rightarrow
{\zeta}^{\kappa_c}(z){\zeta}^{\kappa_c}(z)$ (and $z_{\rm max} = z_{\rm
  dec}$), $C^{\kappa_c\kappa_c}(\ell)$ has the same functional form as
$C_{ij}^{\kappa\kappa}(\ell)$. 
Similarly, the cross-correlation spectrum
$C^{\kappa\kappa_c}_{i}(\ell)$ has the same form as
$C^{\kappa\kappa}_{ij}(\ell)$ with the substitution
${\zeta}_j^{\kappa}(z) \rightarrow {\zeta}^{\kappa_c}(z)$, and we
obtain $C_{i}^{g{\kappa_c}}(\ell)$ by replacing
${\zeta}_j^{\kappa}(z)$ with ${\zeta}^{\kappa_c}(z)$ in
$C_{ij}^{g\kappa}(\ell)$. 

In Figure~\ref{fig:galwl} we show the logarithmic derivatives of the CMB temperature, E-mode, 
and lensing potential power spectra with cosmological parameters. In particular, it is evident 
that both CMB temperature and lensing measurements of the early and late universe are 
necessary to break the degeneracies between early dark energy and sum of neutrino masses.

\subsection{ISW-Lensing Correlation}
\label{iswlen}

We consider five distinct lensing cross-correlations: the first is
that of weak lensing in redshift bins ($i$) with the lensing of the CMB
photons, $C_{i}^{\kappa\kappa_c}(\ell)$; the second is that of galaxy counts
in redshift bins ($i$) with the lensing of the CMB photons, 
$C_{i}^{g\kappa_c}(\ell)$; the third is that of galaxy counts
in redshift bins ($i$) with weak lensing in redshift bins ($j$), $C_{ij}^{g\kappa}(\ell)$; the fourth is 
that of weak lensing in redshift bins ($i$) with the unlensed temperature, $C_{i}^{{\kappa}T}(\ell)$; and 
the fifth is that of lensing of CMB photons with the unlensed temperature, $C^{\kappa_cT}(\ell)$. Having 
already described the first three correlations, we here discuss the last two correlations that arise due to the Integrated
Sachs Wolfe (ISW) effect~\cite{Sachs:1967er, Fosalba:2003ge, Giannantonio:2008zi, Ho:2008bz}.

The perturbations in the CMB temperature due to the late-time
gravitational redshifting of the photons is encapsulated
in~\cite{Hu:2000ee} 
\begin{equation}
{{{\Delta}T^{\rm ISW}({\bf \hat{n}})} \over T} = \int_0^{z_{\rm ls}} dz K(z) {{\zeta}^{T}(z)} {\dot{\Phi}({\bf \hat{n}},z)},
\label{eq:cc10}
\end{equation}
where the overdot denotes a derivative with time $t$, and the ISW weight is 
\begin{equation}
{\zeta}^{T}(z) = {{2} \over {{(1+z)H(z)}}} .
\label{iswt}
\end{equation}
The cross spectrum of the respective Fourier coefficients is given by
\begin{equation}
{1 \over T} {\left\langle{\phi_i^{*}({\bf l})\Delta T({\bf l}^{\prime})}\right\rangle} = (2\pi)^2\delta_{\rm D}({\bf l} - {\bf l}^{\prime}) C_i^{{\phi}T}(\ell) ,
\label{eq:ccextra3}
\end{equation}
which via the Limber approximation reduces to
\begin{equation}
C_i^{{\phi}T}(\ell) = \int_0^{z_{\rm max}} dz K(z) {{\zeta}_i^{\kappa}(z){\zeta}^{T}(z)} {{H(z)} \over {\chi^2(z)}} P_{\Phi\dot{\Phi}}({k},z) .
\label{eq:zz7}
\end{equation} 
To find $P_{\Phi\dot{\Phi}}({k},z)$, the time derivative of the potential
\begin{equation}
\dot{\Phi}({\bf k},z) = {3 \over 2}{{\Omega_m} \over {(1+z)^{-1}}}{\left({H_0 \over k}\right)}^2{\left({{\delta({\bf k},z)} \over {H^{-1}(z)}} - {{\dot{\delta}({\bf k},z)}}\right)}, 
\label{eq:cc4}
\end{equation}
and hence 
\begin{equation}
P_{\Phi\dot{\Phi}}({k},z) = {9 \over 4} {\Omega_m^2 \over {(1+z)^{-2}}} {\left({H_0 \over k}\right)}^4 {\left(P_{\delta\dot{\delta}} - {{P_{\delta\delta}} \over {H^{-1}(z)}}\right)} .
\label{eq:cc6}
\end{equation}
To compute $P_{\delta\dot{\delta}}$, we note that in the linear regime
$P_{\delta\dot{\delta}} = {1 \over 2} {{\partial} \over {\partial t}} P_{\delta\delta}$. 
This relation has been numerically verified to hold in the nonlinear regime up to $\ell=5000$
\cite{Nishizawa:2007pv}. Hence, we adopt this relation so that
\begin{equation}
P_{\Phi\dot{\Phi}}({k},z) = -{9 \over 4}{{\Omega_m^2 H(z)} \over {(1+z)^{-2}}}{\left({H_0 \over k}\right)}^4 {{\left(1 + {{1+z} \over 2}{{\partial} \over {\partial z}}\right)}} P_{\delta\delta} , 
\label{eq:cc6a}
\end{equation}
and the resulting $\kappa -T$ angular power spectrum is then given by 
\begin{eqnarray}
\lefteqn{C_i^{{\kappa}T}(\ell) = {{A_{\ell}^{{\kappa}T}}}
  \int_0^{z_{\rm max}} dz K(z) {{\zeta}_i^{\kappa}(z) {\zeta}^{T}(z) \chi^5(z)}~\times}\\ 
& &\quad (1+z)^2 H^2(z) {\left(1 + {{1+z} \over 2}{{\partial} \over {\partial z}}\right)} \Delta^2_{\delta\delta}\left({{{\ell+1/2}\over{\chi(z)}},z}\right) ,\nonumber 
\label{eq:cc8}
\end{eqnarray}
where the amplitude $A_{\ell}^{{\kappa}T} = \left({{\ell(\ell+1)}/2}\right)^{-1} A_{\ell}^{\kappa\kappa} = {{(9{\pi}^2/4)\left({\Omega_m H_0^2}\right)^2\ell(\ell+1)}{(\ell+1/2)^{-7}}}$. As the derivative $\left|{{{1+z} \over 2}{{\partial} \over {\partial z}}}\Delta^2_{\delta\delta}\right| < \Delta^2_{\delta\delta}$, and the lensing weight contains a negative sign, it follows that $C_i^{{\kappa}T}(\ell) < 0$. We obtain $C^{{\kappa_c}T}(\ell)$ by replacing ${\zeta}_i^{\kappa}(z)$ with ${\zeta}^{\kappa_c}(z)$ in the above equation.

\subsection{ISW-Galaxy Correlation}

To obtain the ISW-galaxy correlation, 
\begin{equation}
C_i^{gT}(\ell) = \int_0^{z_{\rm max}} dz K(z)
{{\zeta}_i^{g}(z){\zeta}^{T}(z)} {{H(z)} \over {\chi^2(z)}}
P_{g\dot{\Phi}}({k},z) , 
\label{eq:zz8}
\end{equation} 
we follow the same steps as for the ISW-lensing correlation and obtain 
\begin{eqnarray}
&&{P_{g\dot{\Phi}}({k},z) = {3 \over 2} {\Omega_m \over {a(z)}} {\left({H_0 \over k}\right)}^2 b(k,z) r_c(k,z)}\\ 
&&\hspace{1.5cm} \times~{\left(H(z)P_{\delta\delta}({k},z) - P_{\delta\dot{\delta}}({k},z)\right)} . \nonumber
\label{eq:cc888}
\end{eqnarray}
We thereby find the projected spectrum
\begin{eqnarray}
\lefteqn{C_i^{{g}T}(\ell) = {{A_{\ell}^{{g}T}}} \int_0^{z_{\rm max}} dz K(z) {{\zeta}_i^{g}(z) {\zeta}^{T}(z)}} \\
&&\hspace*{1.2cm} \times ~ (1+z) H^2(z) \chi^3(z) b(k,z) r_c(k,z) \nonumber \\
&&\hspace*{1.2cm} \times ~ {\left(1 + {{1+z} \over 2}{{\partial} \over {\partial z}}\right)} \Delta^2_{\delta\delta}\left({{{\ell+1/2}\over{\chi(z)}},z}\right) , \nonumber
\label{eq:cc777}
\end{eqnarray}
where amplitude $A_{\ell}^{{g}T} = -{{3{\pi}^2}({\Omega_m H_0^2}){\left({\ell+1/2}\right)^{-5}}}$.

\subsection{Supernovae}
\label{snlabel}

We include the supernovae distances in our analysis for their ability
to constrain the expansion history of the low-redshift universe~\cite{Sarkar}. The
observable quantity is the apparent magnitude:  
\begin{equation}
m = 5\log_{10}{\left({{(1+z)\chi(z)} \over {\rm Mpc}}\right)} + 25 + M,
\label{eqsne}
\end{equation}
where $M$ is
the absolute magnitude, and we marginalize over it around the fiducial
value of $M_{\rm fid} = -19.3$ with a Gaussian prior of 0.6. This is
equivalent to treating the distance modulus $\mu(z) = m(z) - M$ as the
observable, and adding to that the nuisance parameter ${\mathcal
  M}$. The fisher matrix of the prospective supernova
measurements is then   
\begin{equation}
F^{\rm SN}_{\alpha\beta} = {1 \over {\sigma_m^2}} \int_0^{z_{\rm max}}{dz N(z) {{\partial m} \over {\partial p_\alpha}} {{\partial m} \over {\partial p_\beta}}},
\label{snfish}
\end{equation}
where $\sigma_m$ is the measurement uncertainty, $N(z)$ is the
redshift distribution of the SNe, $z_{\rm max}$ is the redshift of the
farthest SN, and $p$ represents the cosmological parameters. 

Although, we do not expect a strong measurement of EDE from SNe, we
will see that it does help in breaking some of the degeneracies in the
parameters of the background cosmology, in particular the dark energy
at the present epoch. 
To this end, we uniformly distribute a set of 300 SNe at
$z<0.1$~\cite{Aldering2002, Albrecht}. For the space-based JDEM
survey, we add 2000 SNe in the range $0.1<z<1.7$~\cite{Aldering},
whereas for the LSST survey we distribute $3.0 \times 10^5$ SNe
(corresponding to six years of data) between
$0.1<z<0.8$~\cite{lsstbb}.   
For each supernova we take the intrinsic noise to be Gaussian in
magnitude with $\sigma_{\rm int} = 0.1$~\cite{KimS}. We divide the
Hubble diagram for $z>0.1$ into 50 redshift bins, and associate each
bin with a redshift-dependent systematic error floor of magnitude   
\begin{equation}
\delta_m(z) = 0.01(0.1/{\Delta z})^{1/2}(1.7/z_{\rm max})(1+z)/2.7, 
\label{snmagerr}
\end{equation}
where $\Delta z$ is the width of the relevant redshift bin~\cite{LinderS}. 
For the ground-based LSST survey, we allow for photometric
uncertainties of the form  
\begin{equation}
\sigma_z = {\sigma_F (1+z) \over {\sqrt{N_c}}}, 
\label{sigzeqn}
\end{equation}
where $N_c = 100$ is the number of spectra used in the photo-z
calibration in each bin, and $\sigma_F = 0.01$. Assuming no
correlation between bins, the total magnitude uncertainty is thereby
quantified via  
\begin{equation}
\sigma_m(z) = \sqrt{{1 \over N_{\rm bin}}\sigma_{\rm int}^2 + {\left({{\partial m} \over {\partial z}} \right)}^2 \sigma_z^2 + \delta_m^2(z)},
\label{snerr}
\end{equation}
where $N_{\rm bin}$ is the number of SNe in each redshift bin, and
$\sigma_z = 0$ for the spectroscopic JDEM survey. The dominant
contribution to the total error comes from the systematic floor, and
the photometric redshift errors turn significant only for more
pessimistic choices of $\sigma_F$. 

For a fixed $z_{\rm max}$, the SN constraints on the underlying
cosmology are robust to variations in the particular SN redshift
distribution of the survey, while this maximum redshift itself imposes
an important limiting factor on the cosmological
constraints~\cite{Frieman:2002wi, Huterer:2000mj, Kim:2003mq,
  LinderS}. Due to the systematic floor, we find the LSST constraints
on the underlying cosmology from SNe remain effectively unchanged
after the first year's data. The existence of a systematic floor
combined with SNe extending to higher redshifts explain why our JDEM
configuration provides stronger constraints than our LSST
configuration when it comes to SN explorations of dark energy (as
shown in Table~\ref{table:withoutomk}).  

\section{Results}

We next explore the constraints on early dark energy from future weak
lensing, large-scale structure, SN, and CMB measurements (and all
relevant cross-correlations). We begin by listing the observational
properties of these probes. Then we elaborate on the constraints on 
the underlying cosmology of the universe, and the biases in the
cosmological parameters that arise due to a potential neglect of
early dark energy. 

\subsection{Survey Properties}
\label{subsa}

We consider Planck~\cite{planckbb, plancksite} for CMB temperature, 
polarization, and lensing measurements, and compare this to a future 
mid-cost CMBPol~\cite{Bock:2009xw, Baumann:2008aq} mission (Epic-2m).
For the large-scale structure, weak lensing, and SN observations, we 
compare two generic types of surveys: 1) a "wide" survey, and 2) a "deep" survey.

For the "wide" configuration, we consider a survey covering half of the sky, such as 
the ground-based Large Synoptic Survey Telescope (LSST)~\cite{Zhan,LSSTsite}.
For comparison, for the "deep" configuration, we consider a survey covering a tenth of 
the sky, such as a space-based Joint Dark Energy Mission (JDEM) candidate like the
SuperNova Acceleration Probe (SNAP)~\cite{SNAPsite, jdemsite, Aldering, RefregSNAP}.
For weak lensing and galaxy clustering measurements, beyond the sky coverage, we only allow for the number density of source galaxies to be different between these two survey configurations (as seen in Table~\ref{table:specslss}). For the SNe, the two survey configurations are different in the number and redshift extent of observations (as seen in Sec.~\ref{snlabel}).
We refer to the wide survey as LSST, and the deep survey as JDEM. Other surveys with similar characteristics include WFIRST~\cite{wfirstsite,wfirstreport} and Euclid~\cite{euclidsite,euclidreport}.

The distribution of source galaxies is divided into five tomographic
redshift bins, as the gain in cosmological information diminishes
rapidly for more than five bins~\cite{MaHu, Joudaki}. The number
density of sources in a square arcminute in each redshift bin (with
boundaries $z_i<z_s<z_{i+1}$) is defined by  
\begin{equation}
\bar{n}_i = \int_{z_i}^{z_{i+1}}\! dz_s\,\rho (z_s), 
\label{nieqn}
\end{equation}
where the redshift distribution of source galaxies is taken to be of
the form~\cite{Wittman}: 
\begin{equation}
\rho (z_s) = \bar{n}_g {z^{\alpha} \over 2 z_0^3} e^{-(z_s/z_0)^{\beta}}.
\label{roeqn}
\end{equation}
We adopt $\left\{z_0=0.5,\ \alpha=2,\ \beta=1 \right\}$~\cite{Zhan}, appropriate 
for LSST~\cite{LSSTsite, Ivezic}, normalized such that
$\int_0^\infty dz\,\rho (z) = \bar{n}_g$. We use the same distribution
to describe our JDEM source population \cite{SNAPsite, Aldering,
  RefregSNAP} with a factor of 2 higher $\bar{n}_g$ value (see
Table~~\ref{table:specslss}). 

\begin{table}[!t]\footnotesize
{\sc Large-Scale Structure Survey Properties}
\begin{tabular}{lcccccccccc|c}
\hline
Probe  & $f_{\rm {sky}}$ & $\bar{n}_{\rm g}$ ({arcmin}$^{-2}$) & $z_{\rm {peak}}$ & $\sqrt{\left\langle \gamma^2 \right\rangle}$ & $\ell_{\rm max}$ & No. bins\\
LSST            &  0.5       & 50   & 1.0   &  0.22 &  3000 & 5 \\
JDEM            &  0.1       & 100  & 1.0   &  0.22 &  3000 & 5 \\
LSST$_{1000}$   &  0.5       & 50   & 1.0   &  0.22 &  1000 & 5 \\
JDEM$_{1000}$   &  0.1       & 100  & 1.0   &  0.22 &  1000 & 5 \\
\hline
\end{tabular}
\caption{Description of the ground-based (LSST) and space-based (JDEM) probes. The redshift 
bins of the source distribution are delimited at [$0.75$, $1.1$, $1.45$, $1.95$, $3.0$], such that each tomographic bin contains 
roughly the same number density of galaxies. The minimum angular multipole $\ell_{\rm min} = 2$ (see text for more details).}
\label{table:specslss}
\end{table}

The observed convergence power spectrum, identical to that of the shear~\cite{Schneider}, is contaminated
by shot noise due to the finite source density, as well as uncertainty in the
intrinsic shapes of the source galaxies, leading to
\begin{equation}
\tilde{C}^{\kappa\kappa}_{ij}(\ell) = \sqrt{{2f^{-1}_{{\rm sky};\kappa}} \over {{2\ell+1}}} \left({C^{\kappa\kappa}_{ij}(\ell) + \delta_{ij} \left\langle \gamma^2 \right\rangle /\bar{n}_i}\right) ,
\label{eqtilde}
\end{equation}
which assumes that the noise is uncorrelated between
tomographic bins. Similarly, for the galaxy angular power spectrum,
\begin{equation}
\tilde{C}^{gg}_{ij}(\ell) = \sqrt{{2f^{-1}_{{\rm sky};g}} \over {{2\ell+1}}} \left({C^{gg}_{ij}(\ell) + \delta_{ij}/\bar{n}_i}\right) .
\end{equation}
We take the intrinsic shape uncertainty of the source galaxies to
be redshift independent: $\left\langle \gamma^2 \right\rangle ^{1/2} = 0.22$, in
accordance with expected results for the future ground-based probe. For simplicity,
we keep the same source distribution and intrinsic shear uncertainty for the surveys on 
ground and space, modifying only the source density to twice that of the ground-based survey,
and the width of the survey to a tenth of the sky. Table~\ref{table:specslss} summarizes the 
characteristics of the two surveys.

\begin{figure}[!t]
\epsfxsize=3.4in
\vspace{-0.8em}
\centerline{\epsfbox{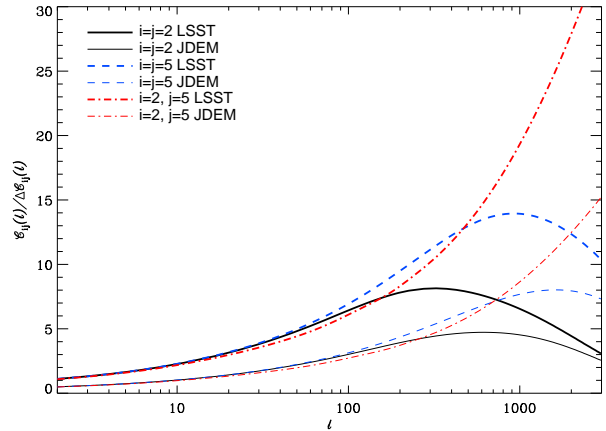}}
\vspace{-1.5em}
\caption{Weak lensing signal-to-noise $C^{\kappa\kappa}_{ij}(\ell)/{\Delta{C^{\kappa\kappa}_{ij}(\ell)}}$ for combinations of the second and fifth tomographic bins in the fiducial cosmology, for LSST (thick) and JDEM (thin).}
\label{fig:wlsigtonoise}
\end{figure}

In Figure~\ref{fig:wlbins}, we plot the power spectrum of the convergence for a subset of the five
tomographic bins (with the redshift divisions listed in the caption of Table~\ref{table:specslss}). We divide the power spectrum by its noise, 
\begin{equation}
{\Delta}C^{\kappa\kappa}_{ij}(\ell)=\sqrt{\left({\tilde{C}^{\kappa\kappa}_{ii}(\ell)\tilde{C}^{\kappa\kappa}_{jj}(\ell)+\tilde{C}^{\kappa\kappa}_{ij}(\ell)^2}\right)/2},  
\end{equation}
in Figure~\ref{fig:wlsigtonoise}. Whereas cosmic variance dominates the error
on large angular scales, the shot noise is dominant on small scales. The signal to noise is consistently higher for the wider LSST 
than it is for the deeper JDEM. The larger width of LSST gives strong signal to noise 
in particular at low-$\ell$, while the greater depth of a JDEM-type survey
makes it increasingly competitive at high-$\ell$.

For the weak lensing surveys we analyze multipoles between $\ell = 2 -
3000$ in one case, and multipoles of $\ell = 2 - 1000$ as a second
case. The cutoffs at $\ell=[1000, 3000]$ largely
avoid non-Gaussianities of the  convergence
field~\cite{Scoccetal,CoorayHu,TakadaJain08}, as well as uncertainties
from  baryonic physics that increase at larger
multipoles~\cite{White,ZhanKnox,vanDaalen:2011xb,Semboloni:2011fe}.
We impose a cutoff in the matter power spectrum for $k > k_{\rm max} =
10~h/{\rm Mpc}$, which implies that for the smallest angular scales,
$\ell_{\rm max} = [1000, 3000]$, distances below $\chi = [100,
300]~{\rm Mpc}/h$ (or equivalently, redshifts below $z \simeq [1/30,
1/10]$) are wiped out in the calculation of the convergence
spectrum. Thus, the cutoffs ensure that scales smaller than $k =
10~h/{\rm Mpc}$ in the matter power spectrum are only probed at low
redshift, where the number densities of sources approach zero. 
Recent hydrodynamic simulations that include AGN feedback to 
solve the overcooling problem have shown these cutoffs may not be conservative
enough unless the baryonic effects in the power spectrum are precisely 
modeled~\cite{vanDaalen:2011xb,Semboloni:2011fe}.

At the other end of the spectrum, we keep all multipoles down to
$\ell_{\rm min} = 2$. We incorporate perturbations in the dark energy
fluid on these scales. 
While a JDEM-type survey is not expected to probe scales below $\ell \sim 10$, we have checked that
our extension down to $\ell = 2$ does not significantly affect our results.
We use the Limber approximation~\cite{Limber, LoVerde:2008re} for all angular scales for
computational reasons. This will lead to some errors in the
constraints but this is not a significant source of concern because
the Limber approximation has been shown to work better than 3\% for
$\ell>20$, better than 5\% for $\ell>10$, and better than 30\% even for 
$\ell>2$ (see Fig.~19 of Ref.~\cite{Smith:2009pn}). 
For the galaxy surveys we eliminate nearly all nonlinear scales in our
analysis by setting $[C_{ij}^{gg}, C_{i}^{gT}, C_{ij}^{g\kappa},
C_{i}^{g\kappa_c}]\left({\ell>\ell_{\rm max}}\right) = 0$. The minimum
angular scale $\ell_{\rm max}$ is computed via $\ell \simeq k \chi$,
where we let $\chi = \chi\left({z_{\rm median}}\right)$ and
approximate, for the redshifts under consideration, $k = k_{\rm
  max}\left({z_{\rm median}}\right) \simeq 0.1 \times
(1+{{z^{3/2}_{\rm median}} / {2}})$ \cite{Rassat:2008ja}. The cutoff
scales are listed in Table~\ref{table:galcutoffs}.  

In the next section (Sec.~\ref{cpf}) we will find, as also reported by
the Dark Energy Task Force~\cite{Albrecht}, weak lensing to be
strongest future probe of the underlying cosmology of the universe,
owing to its sensitivity to both structure formation and the universal
expansion. However, the optimism associated with lensing is predicated
on overcoming the vast systematic uncertainties in both 
measurement and in theory~\cite{Huterer, Hut05, Hirata, Joachimi:2009ez, Bridle:2007ft, Kitching:2008vx,Heymans:2006nu,Schneider:2009es, Scoccetal, HirSel,
MaHu, White, ZhanKnox, Rudd, Hut06, Saito:2008bp, McDonald, Joudaki, CooHu, ShaCoo, Zentner:2007bn}. 
These systematics include dark energy corrections to the modeling of the nonlinear
matter power spectrum~\cite{Huterer, McDonald, Joudaki,Saito:2008bp}, higher order
correction terms in the lensing integral (such as due to the Born
approximation and lens-lens coupling~\cite{CooHu,HirSel,ShaCoo}), and
uncertainties of the matter power spectrum on nonlinear scales due to
the strong influence of baryonic physics~\cite{Hut05, ZhanKnox,White, Rudd, Zentner:2007bn,vanDaalen:2011xb,Semboloni:2011fe}. Observational systematics include
intrinsic galaxy alignments (e.g.~\cite{Hirata, Joachimi:2009ez, Bridle:2007ft, Kitching:2008vx,Heymans:2006nu,Schneider:2009es}),
photometric redshift uncertainties, shear calibration errors, and PSF
anisotropies~\cite{MaHu, Hut06, Kitching:2008vx}. 

\begin{table}[!t]\footnotesize
{\sc Nonlinear Cutoffs in Galaxy Surveys}
\begin{tabular}{lcccccccccc|c}
\hline
Bin  & 1 & 2 & 3 & 4 & 5\\
$z_{\rm median}$   &  0.38 & 0.93 & 1.3 & 1.7 & 2.5 \\
$k_{\rm max}$   & 0.11   &   0.14   &  0.17 &     0.21 &    0.29 \\
$\ell_{\rm max}$   &  120 &      320  &     490   &    720  &     1200 \\
\hline
\end{tabular}
\caption{We effectively eliminate nonlinear scales in the constraints from galaxy surveys by setting $[C_{ij}^{gg}, C_{i}^{gT}, C_{ij}^{g\kappa}, C_{i}^{g\kappa_c}]\left({\ell>\ell_{\rm max}}\right) = 0$. The minimum angular scale $\ell_{\rm max}$ is computed via $\ell+1/2 = k \chi$, where we let $\chi = \chi\left({z_{\rm median}}\right)$ and approximate, for the considered redshifts, $k = k_{\rm max}\left({z_{\rm median}}\right) \simeq 0.1 \times (1+{{z^{3/2}_{\rm median}} / {2}})$ \cite{Rassat:2008ja}.}
\label{table:galcutoffs}
\end{table}

\begin{table}[!t]\footnotesize
{\sc CMB Survey Properties}
\begin{tabular}{lcccccccccc|c}
\hline
Experiment & Channel & FWHM  & $\Delta T/T \times 10^6$ &  $\Delta P/T \times 10^6$ \\
Planck  & 100    & 10    & 25        &  40 \\
 & 143    & 7.1   & 16        &  30 \\
 & 217    & 5.0   & 24        &  49 \\
EPIC-2m  & 100    & 8.0   & 0.84       &  1.19 \\
 & 150    & 5.0   & 0.81       &  1.15 \\
 & 220    & 3.5   & 1.24       &  1.75 \\
\hline
\end{tabular}
\caption{Experimental specifications for the Planck and mid-cost CMBPol (EPIC-2m) missions. 
The sky fraction $f_{\rm sky} = 0.65$, and the angular multipoles extend from $\ell_{\rm min} = 2$ to $\ell_{\rm max}=2000$. The channel frequencies are given in GHz, and the angular resolutions in arcminutes.}
\label{table:cmbspecs}
\end{table}

Furthermore, the observed ellipticities of weakly lensed galaxies are sensitive
to the reduced shear, $g = \gamma / (1-\kappa)$, where $\gamma$ is the shear and
$\kappa$ is the convergence. In the weak lensing regime we make use of
the expansion of the reduced shear to first order in the fields: $g \approx
\gamma$. For future lensing surveys, it has been
shown that this approximation induces a bias on the cosmological
parameters at the same order as that of the parameter 
constraints~\cite{Shapiro:2008yk,DodShaWhi}. Our main motivation is to
elucidate the degradation in constraints due to EDE and the biases in
the cosmological parameters from neglect of EDE. We therefore
continue with the above mentioned assumption of the shear as the
lensing observable.  

Intrinsic galaxy alignments reflect the reality that underlying galaxies that become lensed are not inherently circularly shaped, but possess a non-trivial intrinsic ellipticity (e.g.~\cite{Hirata, Joachimi:2009ez, Bridle:2007ft, Kitching:2008vx,Heymans:2006nu,Schneider:2009es}).
The assumption that these intrinsic ellipticities are uncorrelated with each other and the gravitational shear breaks down for high-precision measurements of weak lensing from next-generation experiments. 
For instance, nearby galaxies that lie in the same gravitational potential of a massive dark matter halo will experience the same tidal forces, giving
rise to an intrinsic-intrinsic term~\cite{Hirata, Joachimi:2009ez, Bridle:2007ft, Kitching:2008vx,Heymans:2006nu,Schneider:2009es}. 
Moreover, intrinsic-shear correlations occur for pairs of galaxies where one galaxy is correlated with the surrounding density field in the foreground, thereby contributing to the lensing distortion of the background galaxy~\cite{Hirata,Joachimi:2009ez, Bridle:2007ft, Kitching:2008vx,Heymans:2006nu,Schneider:2009es}.

While intrinsic ellipticity correlations can be controlled by downweighting close galaxy pairs from the analysis~\cite{Heymans:2003hx,King:2002tp,King:2002fc}, the removal of shear-ellipticity correlations is more difficult~\cite{Hirata,Heymans:2006nu,Joachimi:2009ez, Bridle:2007ft,Kitching:2008vx,Schneider:2009es,Zhang:2008pw,Joachimi:2008ea,King:2005ng}. 
Ref.~\cite{Bridle:2007ft} has shown that the dark energy EOS may be biased by 50\% if intrinsic alignments are not accounted for in next-generation weak lensing experiments. 
The presence of intrinsic alignments also tighten the requirements on photometric redshift uncertainties by a factor of three, regardless of the fraction of catastrophic outliers~\cite{Bridle:2007ft}. 
Suggested approaches to avoid these degradations include halo modeling~\cite{Schneider:2009es} and nulling techniques~\cite{Joachimi:2008ea}. 
Alternatively, the systematic may be self-calibrated for an extended set of observables that can correlate with the weak lensing shear and intrinsic alignment signals, such as galaxy clustering and lensing magnification~\cite{Zhang:2008pw,Joachimi:2009ez} (also see~\cite{Kitching:2008vx}).

The effects of multiplicative and additive uncertainties in the
convergence (e.g. from shear miscalibration and PSF anisotropies,
respectively) on the measurement of dark energy are influenced by the
corresponding priors. As shown in Ref.~\cite{Hut06} (also see
Refs.~\cite{Joudaki, TakadaJain08}), the cosmological parameter
constraints are compromised for multiplicative systematics at the 1\%
level, and mean additive shear systematics at the $10^{-5}$ level. 
The situation is analogous for the uncertainty in the photometric
redshift distribution of the sources, where the parameter constraints
from lensing are either heavily influenced ($\gsim1\%$ prior) or
minimally influenced ($\lsim0.1\%$ prior) by the photometric
uncertainties~\cite{MaHu, Hut06, Joudaki}. Fortunately, it has been
shown that a complementary spectroscopic sample of $10^4-10^5$
galaxies efficiently protects against photometric redshift errors as
well as catastrophic outliers~\cite{Amara:2006kp}, whereas alternative
methods may even satisfy the systematic requirements from photometry
alone~\cite{Bordoloi:2009wb, Quadri:2009ar}. 

Thus, in this work, we will assume that these systematic difficulties
have been largely overcome with minimal influence on the constraints
by the time the data from the considered next-generation lensing
probes are analyzed. At the same time, we are not incorporating
further statistics that can be extracted from weak lensing, such as
that included in the bispectrum~\cite{Bernardeau:1996un,Hui:1999ak,Cooray:2000uu,TakadaJain}, or utilizing the
complementarity between measurements of shear and
magnification~\cite{VanWaerbeke:2009fb,Joachimi:2009ez}.

We end this section with a summary of the CMB temperature,
polarization and lensing noise properties. The effective experimental
noise power spectrum associated with the temperature and polarization
fields is given by a summation over the number of channels,  
\begin{eqnarray}
N^{aa}(\ell)=\left[\sum_{i=1}^{N_{\rm chan}}\left(\left({\Delta a \over T}\right)_ie^{\el(\el+1)\theta_i/16\ln 2}\right)^{-2}\right]^{-1},
\label{eqn:pnoisepower}
\end{eqnarray}
where $\Delta a$ is the detector noise for $a \in (T,E)$, $\theta$
denotes the beam FWHM,  and we assume $N^{TE}(\ell)=0$. 
The optimal noise power spectrum of a quadratic estimator of the convergence 
field is given by \cite{Hu:2001fa,Okamoto:2003zw}
\begin{eqnarray}
N^{\kappa_c\kappa_c}(\ell) &=&
	\left[ \sum_{\el_1 \el_2} {
        ( C_{\el_2}^{TT} F_{\el_1 \el \el_2} +
          C_{\el_1}^{TT} F_{\el_2 \el \el_1} )^2
          \over 2 (C_{\el_1}^{\tilde{T}\tilde{T}} + N_{\el_1}^{\tilde{T}\tilde{T}})
		  (C_{\el_2}^{\tilde{T}\tilde{T}} + N_{\el_2}^{\tilde{T}\tilde{T}}) }\right]^{-1}\nonumber\\
&&\times
	~(\el(\el+1)/2)^2 (2\el+1) ,
	\label{eqn:deflectionnoise}
\end{eqnarray}
where $\tilde{T}$ denotes the lensed temperature, and
\begin{eqnarray}
F_{\el_1 \el \el_2}& =& \sqrt{ (2\el_1+1)(2\el+1)(2\el_2+1) \over 4\pi} 
\wj{\el_1}{\el}{\el_2}{0}{0}{0}
\nonumber\\
&& \times ~{1 \over 2} [\el(\el+1)+\el_2(\el_2+1)-\el_1(\el_1+1)] ,
\end{eqnarray}
where the quantity in brackets is the Wigner-3j symbol. Finally, we define
\begin{equation}
\tilde{C}^{ab}(\ell)=\sqrt{{2f^{-1}_{{\rm sky;cmb}}} \over {{2\ell+1}}} \left({C^{ab}(\ell)+\delta_{ab}{N^{ab}(\ell)}}\right) , 
\end{equation}
where $\left\{a,b\right\} \in \left\{T,E,\kappa_c\right\}$. Values for
the considered CMB experiments are given in
Table~\ref{table:cmbspecs}. Secondary non-Gaussianities in the covariance
from the trispectrum (due to weak lensing, the ISW effect, and the SZ
effect) have been shown to degrade the Planck and EPIC parameter
constraints by 20\% and 30\% \cite{Smith:2006nk,Smidt:2009qa}
respectively; however, their full account lies beyond the scope of 
this work.  

\begin{figure*}[htbp]
\vspace{-0.6em}
\begin{center}
\leavevmode
\includegraphics[scale=0.47]{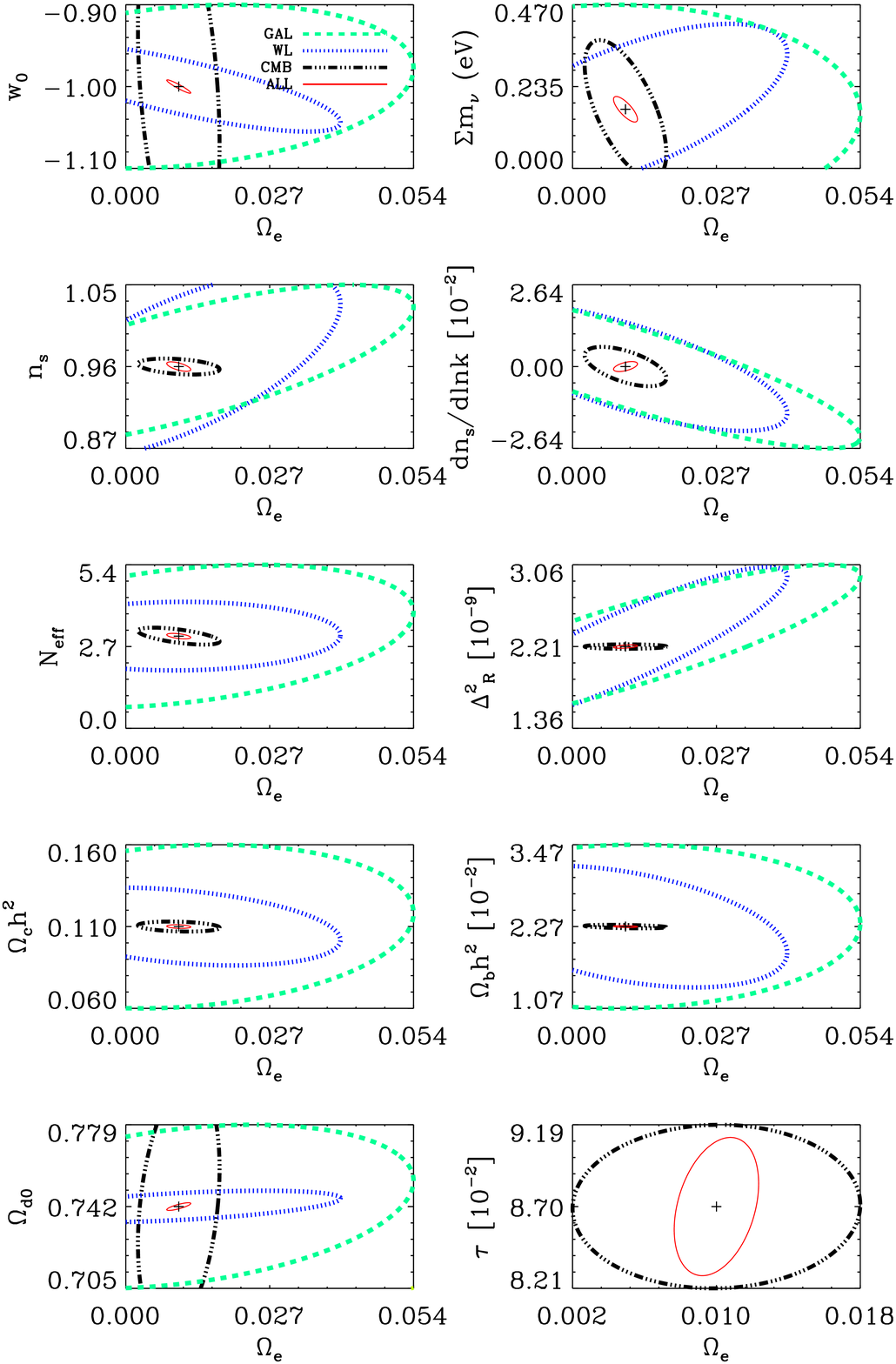}
\end{center}
\vspace{-2.1em}
\caption{Parameter degeneracies with early dark energy density
  ($\Omega_e$) in a flat 
  universe (also see Table~\ref{table:withoutomk}) for Planck measurements of
  temperature and CMB 
  lensing spectra [TT, EE, TE, $\kappa_c\kappa_c$, $\kappa_c$T]
  (dot-dashed, black), along with LSST tomographic weak lensing
  spectra [$\kappa\kappa$] (dotted, blue) and tomographic galaxy spectra
  [gg] (dashed, turquoise). Constraints from SNe
  are too weak to be visible in the shown parameter
  regions. The error ellipses from the combination of all these probes
  (including SNe), incorporating all cross-correlations (see
  Eqn.~\ref{eq:clall}) is shown as (solid, red) curves.}  
\label{fig:ellipses}
\end{figure*}

\subsection{Comprehensive Parameter Forecasts}
\label{cpf}

In previous sections we explored the qualitative influence of EDE on
the lensing, galaxy, supernova, and CMB observables, via its impact on
the expansion rate and matter power spectrum. We now examine how these
corrections quantitatively affect the combined constraints of the dark
energy. To this end, we utilize a Fisher matrix
formalism~\cite{HuJain,Tegmark}: 
\begin{equation}
F^{\rm total}_{\alpha\beta} = \sum_\ell \Delta\ell \times \Tr \left
[{{\bf \tilde C}}_\ell^{-1} {{\partial {\bf C}_\ell} \over {\partial p_\alpha}} {{\bf \tilde C}}_\ell^{-1}
{{\partial {\bf C}_\ell} \over {\partial p_\beta}} \right ] + F^{\rm SN}_{\alpha\beta},
\label{eq:fisher}
\end{equation}
where the decoupled SN fisher matrix is defined in Eqn.~\ref{snfish},
and for the combined observational analysis the symmetric matrix 
\begin{equation}
{\bf C_\ell} = \begin{pmatrix}
C_\ell^{\left\{\kappa\right\}\left\{\kappa\right\}} & C_\ell^{\left\{\kappa\right\}\kappa_c} & C_\ell^{\left\{\kappa\right\} T} & 0 & C_\ell^{\left\{\kappa\right\} {\left\{g\right\}}} \\
C_\ell^{\kappa_c\left\{\kappa\right\}} & C_\ell^{\kappa_c\kappa_c} & C_\ell^{\kappa_c T} & 0 & C_\ell^{\kappa_c\left\{g\right\}}\\
C_\ell^{T\left\{\kappa\right\}} & C_\ell^{T\kappa_c} & C_\ell^{TT} & C_\ell^{TE} & C_\ell^{T\left\{g\right\}} \\
0 & 0 & C_\ell^{ET} & C_\ell^{EE} & 0 \\
C_\ell^{{\left\{g\right\}}\left\{\kappa\right\}} & C_\ell^{\left\{g\right\}\kappa_c} & C_\ell^{\left\{g\right\}T} & 0 & C_\ell^{\left\{g\right\}\left\{g\right\}}
\end{pmatrix} ,
\label{eq:clall}
\end{equation}
such that $\left\{\kappa\right\}$ consists of the spectra from five tomographic 
bins $\left(\kappa_1, \kappa_2, \kappa_3, \kappa_4, \kappa_5\right)$
and $\left\{g\right\}$ consists of the spectra from five tomographic  
bins $\left(g_1, g_2, g_3, g_4, g_5\right)$.
$C_\ell^{\left\{\kappa\right\}\left\{\kappa\right\}}$,
$C_\ell^{\left\{g\right\}\left\{g\right\}}$,
$C_\ell^{\left\{\kappa\right\}\left\{g\right\}}$ are therefore
$5\times5$ submatrices, and $C_\ell^{\left\{\kappa\right\}\kappa_c}$,
$C_\ell^{\left\{\kappa\right\}T}$,
$C_\ell^{\left\{g\right\}\kappa_c}$, $C_\ell^{\left\{g\right\}T}$ are
$5\times1$ submatrices. 
For the terms in Eqn.~\ref{eq:fisher} we carry out two-sided numerical derivatives
with steps of $2\%$ in most parameter values. We have confirmed the
robustness of our results to other choices of step size.

In Tables~\ref{table:withoutomk}-\ref{table:epic}, we illustrate prospective
constraints from Planck/EPIC CMB temperature ($T$), E-mode
polarization ($E$), lensing ($\kappa_c$), LSST/JDEM weak lensing
tomography ($\kappa$), galaxy tomography ($g$), SNe ($s$), and their
combined impact (including all relevant cross-correlations shown in
Eqn.~\ref{eq:clall}) on the 12 considered cosmological parameters
($\Omega_{d0}$, $\Omega_e$, ${\Omega_c}h^2$, ${\Omega_b}h^2$, $\Omega_k$,
$\sum{m_{\nu}}$, $N_{\rm eff}$, $w_0$, $n_s$, ${\rm d}{n_s}/{\rm
  d}\ln{k}$, $\Delta_R^2$, $\tau$).  

The contents of our tables are as follows: In Table~\ref{table:withoutomk} and
Table~\ref{table:withoutomkandede} we consider only a flat universe, with curvature
always considered in the other tables. These tables present the
separate constraints on the underlying cosmology obtained from the
CMB, lensing tomography, galaxy tomography, and SNe, along with the
synergies attained from a combined analysis of these
probes. Table~\ref{table:withoutomkandede} differs from Table~\ref{table:withoutomk} in that
it fixes the early dark energy density. Table~\ref{table:withomk}
differs from Table~\ref{table:withoutomk} in that it allows for variation in
curvature.   
In Table~\ref{table:epic}, we present results where the CMB constraints are
derived from a future experiment like the proposed 2m
EPIC~\cite{Baumann:2008aq, Bock:2009xw} (compare to Table~\ref{table:withomk}). Table~\ref{table:nosnova} differs 
from Table~\ref{table:withomk} in its neglect of SNe measurements. 
Lastly, Table~\ref{table:nocross} differs from Table~\ref{table:withomk}
in that we neglect cross-correlations between the observables
(i.e. neglecting {\it all} correlations between $T$, $E$, $\kappa_c$,
$\kappa$, $g$, except for tomographic cross-correlations within
$\kappa$).


\begin{table*}[th]\footnotesize
\begin{center}
\begin{tabular}{lcccccccccccccc|c}
\hline
\hline
Probe & $w_0$ & $\Omega_{d0}$ & $\Omega_e$ & $\sum{m_{\nu}} \rm{(eV)}$ & $n_s$ & ${dn_s \over d\ln k}$ & $10^{10}\Delta_R^2$ & $\Omega_ch^2$ & $10^3\Omega_bh^2$ & $N_{\rm eff}$ & $\tau$ & $\Omega_k$\cr
P &      0.42 &       0.12 &     0.0086 &       0.47 &     0.0088 &     0.0085 &       0.26 &     0.0032 &       0.27 &       0.28 &     0.0056 & --- \cr
PK &      0.23 &      0.057 &     0.0077 &       0.20 &     0.0085 &     0.0081 &       0.23 &     0.0031 &       0.25 &       0.28 &     0.0050 & --- \cr
P$_{\ell>30}{\rm K}+\sigma(\tau)$&     0.23 &      0.060 &     0.0080 &       0.21 &     0.0085 &     0.0082 &       0.44 &     0.0031 &       0.25 &       0.28 &     0.0097 & --- \cr
L$^{\kappa}$ &     0.055 &     0.0071 &      0.031 &       0.25 &       0.10 &      0.027 &        8.3 &      0.024 &        9.0 &        1.1 & --- & --- \cr
J$^{\kappa}$ &     0.088 &      0.012 &      0.049 &       0.38 &       0.15 &      0.040 &        13. &      0.035 &        14. &        1.7 & --- & --- \cr
L$^{\kappa}_{\ell<1000}$ &     0.072 &      0.013 &      0.081 &       0.38 &       0.17 &      0.052 &        17. &      0.049 &        16. &        1.6 & --- & --- \cr
J$^{\kappa}_{\ell<1000}$ &      0.13 &      0.024 &       0.14 &       0.67 &       0.27 &      0.088 &        28. &      0.077 &        26. &        2.5 & --- & --- \cr
L$^g$ &     0.099 &      0.037 &      0.044 &       0.30 &      0.087 &      0.034 &        8.5 &      0.050 &        12. &        2.4 & --- & --- \cr
J$^g$ &      0.21 &      0.079 &      0.078 &       0.63 &       0.16 &      0.061 &        13. &       0.11 &        26. &        5.2 & --- & --- \cr
L$^s$ &      0.21 &       0.27 &       0.59 & --- & --- & --- & --- &       0.16 & --- & --- & --- & --- \cr
J$^s$ &     0.045 &      0.085 &       0.27 & --- & --- & --- & --- &      0.086 & --- & --- & --- & --- \cr
PKL$^{\kappa}$L$^g$L$^s$ &    0.0081 &     0.0017 &     0.0023 &      0.037 &     0.0050 &     0.0020 &       0.17 &     0.0011 &       0.13 &      0.090 &     0.0042 & --- \cr
PKJ$^{\kappa}$J$^g$J$^s$ &    0.0076 &     0.0014 &     0.0024 &      0.036 &     0.0052 &     0.0023 &       0.18 &     0.0012 &       0.14 &      0.088 &     0.0044 & --- \cr
PKL$^{\kappa}_{\ell<1000}$L$^g_{\ell<1000}$L$^s$ &    0.0084 &     0.0018 &     0.0023 &      0.039 &     0.0054 &     0.0027 &       0.18 &     0.0012 &       0.14 &      0.099 &     0.0043 & --- \cr
PKJ$^{\kappa}_{\ell<1000}$J$^g_{\ell<1000}$J$^s$  &    0.0078 &     0.0015 &     0.0024 &      0.037 &     0.0054 &     0.0035 &       0.19 &     0.0014 &       0.14 &       0.10 &     0.0045 & --- \cr
PKL$^{\kappa}$L$^g$L$^s+\Omega_k$ &    0.0085 &     0.0019 &     0.0023 &      0.038 &     0.0051 &     0.0021 &       0.20 &     0.0013 &       0.13 &      0.090 &     0.0044 &    0.00056 \cr
PKJ$^{\kappa}$J$^g$J$^s+\Omega_k$ &    0.0076 &     0.0021 &     0.0025 &      0.036 &     0.0055 &     0.0024 &       0.19 &     0.0012 &       0.14 &      0.090 &     0.0045 &    0.00075 \cr
\hline
\hline
\end{tabular}
\caption{1$\sigma$ uncertainties on cosmological parameters from a
  combination of probes. P denotes CMB T, E, TE modes for a Planck
  survey. K denotes the CMB lensing potential power spectrum and the
  correlation with the temperature field for Planck. L denotes an LSST
  type survey, whereas J denotes a JDEM type survey, and the
  superscripts $\kappa$, $g$, $s$, refer to weak lensing tomography,
  galaxy tomography, and supernova measurements, respectively. When we
  combine more than one probe, all relevant cross-correlations between
  the selected probes are included. Thus, for the case of
  PKL$^{\kappa}$L$^g$, all cross-correlations between [$T, E,
  \kappa_c, \kappa, g$] are included (see Eqn.~\ref{eq:clall}). The
  subscripts with $\ell<1000$ refer to cutoffs of the respective
  auto-correlations (and all related cross-correlations) at $\ell =
  1000$. At redshifts $z = [0, 1, 2, 3]$ the early dark energy
  constitutes $[0, 2.1, 8.0, 17.7]$\% of the overall amount of dark
  energy (quantified as ${\Omega_d(z)-\Omega_w(z)} /
  {\Omega_d(z)}$ with $w = -1$ and $\Omega_e = 0.01$). For the case
  where $\ell<30$ modes in CMB polarization data are
  excluded, we add a prior of 0.01 on the optical depth.} 
\label{table:withoutomk}
\end{center}
\end{table*}

\begin{table*}[th]\footnotesize
\begin{center}
\begin{tabular}{lcccccccccccccc|c}
\hline
\hline
Probe & $w_0$ & $\Omega_{d0}$ & $\Omega_e$ & $\sum{m_{\nu}} \rm{(eV)}$ & $n_s$ & ${dn_s \over d\ln k}$ & $10^{10}\Delta_R^2$ & $\Omega_ch^2$ & $10^3\Omega_bh^2$ & $N_{\rm eff}$ & $\tau$ & $\Omega_k$\cr
${\rm P}-\Omega_e$  &      0.42 &       0.11 & --- &       0.46 &     0.0086 &     0.0072 &       0.26 &     0.0032 &       0.25 &       0.24 &     0.0055 & --- \cr
${\rm PK}-\Omega_e$ &      0.21 &      0.055 & --- &       0.15 &     0.0079 &     0.0066 &       0.23 &     0.0030 &       0.23 &       0.22 &     0.0050 & --- \cr
P$_{\ell>30}{\rm K}-\Omega_e+\sigma(\tau)$ &     0.22 &      0.058 & --- &       0.17 &     0.0079 &     0.0066 &       0.44 &     0.0030 &       0.23 &       0.22 &     0.0097 & --- \cr
L$^{\kappa}-\Omega_e$ &     0.033 &     0.0061 & --- &       0.19 &      0.075 &      0.018 &        3.9 &      0.023 &        8.1 &        1.1 & --- & --- \cr
J$^{\kappa}-\Omega_e$ &     0.053 &     0.0097 & --- &       0.30 &       0.12 &      0.029 &        6.1 &      0.034 &        13. &        1.7 & --- & --- \cr
L$^{\kappa}_{\ell<1000}-\Omega_e$ &     0.050 &     0.0075 & --- &       0.37 &       0.13 &      0.025 &        6.7 &      0.049 &        16. &        1.5 & --- & --- \cr
J$^{\kappa}_{\ell<1000}-\Omega_e$ &     0.083 &      0.013 & --- &       0.65 &       0.22 &      0.044 &        12. &      0.076 &        26. &        2.4 & --- & --- \cr
L$^g-\Omega_e$ &     0.096 &      0.035 & --- &       0.30 &      0.060 &      0.017 &        4.4 &      0.050 &        12. &        2.2 & --- & --- \cr
J$^g-\Omega_e$ &      0.21 &      0.067 & --- &       0.60 &       0.13 &      0.038 &        8.0 &       0.11 &        26. &        4.7 & --- & --- \cr
L$^s-\Omega_e$ &      0.13 &      0.057 & --- & --- & --- & --- & --- &      0.080 & --- & --- & --- & --- \cr
J$^s-\Omega_e$ &     0.042 &      0.012 & --- & --- & --- & --- & --- &      0.075 & --- & --- & --- & --- \cr
PKL$^{\kappa}$L$^g$L$^s-\Omega_e$ &    0.0026 &     0.0010 & --- &      0.023 &     0.0039 &     0.0017 &       0.16 &     0.0011 &       0.13 &      0.078 &     0.0040 & --- \cr
PKJ$^{\kappa}$J$^g$J$^s-\Omega_e$ &    0.0031 &    0.00092 & --- &      0.023 &     0.0047 &     0.0023 &       0.17 &     0.0012 &       0.13 &      0.084 &     0.0044 & --- \cr
PKL$^{\kappa}_{\ell<1000}$L$^g_{\ell<1000}$L$^s-\Omega_e$ &    0.0028 &     0.0012 & --- &      0.024 &     0.0041 &     0.0025 &       0.16 &     0.0012 &       0.13 &      0.084 &     0.0041 & --- \cr
PKJ$^{\kappa}_{\ell<1000}$J$^g_{\ell<1000}$J$^s-\Omega_e$ &    0.0034 &    0.00099 & --- &      0.023 &     0.0050 &     0.0035 &       0.17 &     0.0014 &       0.14 &      0.098 &     0.0044 & --- \cr
PKL$^{\kappa}$L$^g$L$^s+\Omega_k-\Omega_e$ &    0.0029 &     0.0014 & --- &      0.028 &     0.0039 &     0.0019 &       0.18 &     0.0012 &       0.13 &      0.079 &     0.0041 &    0.00055 \cr
PKJ$^{\kappa}$J$^g$J$^s+\Omega_k-\Omega_e$ &    0.0037 &     0.0020 & --- &      0.025 &     0.0049 &     0.0023 &       0.18 &     0.0012 &       0.14 &      0.085 &     0.0044 &    0.00072 \cr
\hline
\hline
\end{tabular}
\caption{Same as Table~\ref{table:withoutomk} (Planck CMB), for an EDE fiducial cosmology with $\Omega_e = 0.01$ kept fixed.}
\label{table:withoutomkandede}
\end{center}
\end{table*}

\begin{table*}[th]\footnotesize
\begin{center}
\begin{tabular}{lcccccccccccccc|c}
\hline
\hline
Probe & $w_0$ & $\Omega_{d0}$ & $\Omega_e$ & $\sum{m_{\nu}} \rm{(eV)}$ & $n_s$ & ${dn_s \over d\ln k}$ & $10^{10}\Delta_R^2$ & $\Omega_ch^2$ & $10^3\Omega_bh^2$ & $N_{\rm eff}$ & $\tau$ & $\Omega_k$\cr
${\rm P}+\Omega_k$ &      0.43 &       0.12 &     0.0087 &       0.53 &     0.0090 &     0.0085 &       0.26 &     0.0032 &       0.27 &       0.29 &     0.0056 &      0.013 \cr
${\rm PK}+\Omega_k$ &      0.26 &      0.089 &     0.0082 &       0.22 &     0.0085 &     0.0081 &       0.24 &     0.0031 &       0.26 &       0.28 &     0.0053 &     0.0073 \cr
P$_{\ell>30}{\rm K}+\Omega_k+\sigma(\tau)$ &      0.27 &      0.092 &     0.0085 &       0.24 &     0.0085 &     0.0082 &       0.44 &     0.0031 &       0.26 &       0.28 &     0.0097 &     0.0075 \cr
L$^{\kappa}+\Omega_k$ &     0.055 &      0.014 &      0.033 &       0.27 &       0.11 &      0.029 &        8.3 &      0.025 &        9.3 &        1.2 & --- &      0.012 \cr
J$^{\kappa}+\Omega_k$ &     0.089 &      0.025 &      0.055 &       0.42 &       0.16 &      0.045 &        13. &      0.037 &        14. &        1.8 & --- &      0.020 \cr
L$^{\kappa}_{\ell<1000}+\Omega_k$ &     0.073 &      0.018 &      0.083 &       0.41 &       0.18 &      0.062 &        18. &      0.050 &        16. &        1.8 & --- &      0.019 \cr
J$^{\kappa}_{\ell<1000}+\Omega_k$ &      0.13 &      0.033 &       0.14 &       0.70 &       0.29 &       0.10 &        30. &      0.077 &        26. &        2.8 & --- &      0.032 \cr
L$^g+\Omega_k$ &      0.20 &      0.040 &      0.045 &       0.34 &      0.093 &      0.035 &        8.5 &      0.053 &        12. &        3.0 & --- &      0.061 \cr
J$^g+\Omega_k$ &      0.39 &      0.088 &      0.087 &       0.67 &       0.20 &      0.073 &        15. &       0.12 &        27. &        6.4 & --- &       0.12 \cr
L$^s+\Omega_k$ &      0.21 &       0.27 &       0.59 & --- & --- & --- & --- &       0.16 & --- & --- & --- &     0.013 \cr
J$^s+\Omega_k$ &     0.046 &      0.086 &       0.27 & --- & --- & --- & --- &      0.086 & --- & --- & --- &     0.013 \cr
PKL$^{\kappa}$L$^g$L$^s+\Omega_k$ &    0.0085 &     0.0019 &     0.0023 &      0.038 &     0.0051 &     0.0021 &       0.20 &     0.0013 &       0.13 &      0.090 &     0.0044 &    0.00056 \cr
PKJ$^{\kappa}$J$^g$J$^s+\Omega_k$ &    0.0076 &     0.0021 &     0.0025 &      0.036 &     0.0055 &     0.0024 &       0.19 &     0.0012 &       0.14 &      0.090 &     0.0045 &    0.00075 \cr
PKL$^{\kappa}_{\ell<1000}$L$^g_{\ell<1000}$L$^s+\Omega_k$ &    0.0090 &     0.0022 &     0.0024 &      0.040 &     0.0054 &     0.0031 &       0.21 &     0.0013 &       0.14 &       0.10 &     0.0044 &    0.00068 \cr
PKJ$^{\kappa}_{\ell<1000}$J$^g_{\ell<1000}$J$^s+\Omega_k$ &    0.0078 &     0.0023 &     0.0025 &      0.037 &     0.0059 &     0.0035 &       0.20 &     0.0014 &       0.15 &       0.11 &     0.0045 &    0.00081 \cr
\hline
\hline
\end{tabular}
\caption{Same as Table~\ref{table:withoutomk} (Planck CMB), allowing for $\Omega_k$ to vary. We include a CMB prior on the curvature for the SN surveys. For the galaxy measurements alone, the bias parameters defined in Sec.~\ref{galpow} as $b(z) = b_0 + {b_1}z$ are constrained to $\sigma\{b_0, b_1\} = \{0.16, 0.19\}$ for LSST and $\sigma\{b_0, b_1\} = \{0.34, 0.43\}$ for JDEM. Our 10\% prior on these bias parameters (i.e. $\sigma\{b_0, b_1\} = \{0.1, 0.08\}$) therefore improves the constraints on cosmology. When all probes are combined, the bias parameters are constrained to $\sigma\{b_0, b_1\} = \{6.7, 3.5\}\times10^{-3}$ for LSST and $\sigma\{b_0, b_1\} = \{7.4, 4.6\}\times10^{-3}$ for JDEM. An imposed 10\% prior is therefore rendered negligible in the multi-probe scenario.}
\label{table:withomk}
\end{center}
\end{table*}

\begin{table*}[th]\footnotesize
\begin{center}
\begin{tabular}{lcccccccccccccc|c}
\hline
\hline
Probe & $w_0$ & $\Omega_{d0}$ & $\Omega_e$ & $\sum{m_{\nu}} \rm{(eV)}$ & $n_s$ & ${dn_s \over d\ln k}$ & $10^{10}\Delta_R^2$ & $\Omega_ch^2$ & $10^3\Omega_bh^2$ & $N_{\rm eff}$ & $\tau$ & $\Omega_k$\cr
PKL$^{\kappa}$L$^g+\Omega_k$ &    0.0096 &     0.0021 &     0.0026 &      0.040 &     0.0053 &     0.0022 &       0.20 &     0.0013 &       0.13 &      0.092 &     0.0044 &    0.00056 \cr
PKJ$^{\kappa}$J$^g+\Omega_k$ &     0.012 &     0.0033 &     0.0032 &      0.047 &     0.0056 &     0.0024 &       0.19 &     0.0013 &       0.14 &      0.092 &     0.0045 &    0.00088 \cr
PKL$^{\kappa}$L$^g+\Omega_k-\Omega_e$ &    0.0029 &     0.0015 & --- &      0.028 &     0.0039 &     0.0019 &       0.18 &     0.0012 &       0.13 &      0.079 &     0.0041 &    0.00056 \cr
PKJ$^{\kappa}$J$^g+\Omega_k-\Omega_e$ &    0.0045 &     0.0025 & --- &      0.026 &     0.0050 &     0.0023 &       0.18 &     0.0013 &       0.14 &      0.085 &     0.0044 &    0.00087 \cr
PKL$^{\kappa}_{\ell<1000}$L$^g_{\ell<1000}+\Omega_k$ &     0.010 &     0.0024 &     0.0027 &      0.042 &     0.0057 &     0.0032 &       0.21 &     0.0013 &       0.14 &       0.10 &     0.0045 &    0.00068 \cr
PKJ$^{\kappa}_{\ell<1000}$J$^g_{\ell<1000}+\Omega_k$ &     0.012 &     0.0037 &     0.0033 &      0.049 &     0.0060 &     0.0036 &       0.20 &     0.0014 &       0.15 &       0.11 &     0.0045 &    0.00098 \cr
PKL$^{\kappa}_{\ell<1000}$L$^g_{\ell<1000}+\Omega_k-\Omega_e$ &    0.0030 &     0.0020 & --- &      0.029 &     0.0041 &     0.0031 &       0.18 &     0.0013 &       0.14 &      0.084 &     0.0042 &    0.00066 \cr
PKJ$^{\kappa}_{\ell<1000}$J$^g_{\ell<1000}+\Omega_k-\Omega_e$ &    0.0049 &     0.0029 & --- &      0.027 &     0.0053 &     0.0035 &       0.19 &     0.0014 &       0.15 &       0.10 &     0.0044 &    0.00097 \cr
\hline
\hline
\end{tabular}
\caption{Same as Table~\ref{table:withomk}, without SNe.}
\label{table:nosnova}
\end{center}
\end{table*}

\begin{table*}[th]\footnotesize
\begin{center}
\begin{tabular}{lcccccccccccccc|c}
\hline
\hline
Probe & $w_0$ & $\Omega_{d0}$ & $\Omega_e$ & $\sum{m_{\nu}} \rm{(eV)}$ & $n_s$ & ${dn_s \over d\ln k}$ & $10^{10}\Delta_R^2$ & $\Omega_ch^2$ & $10^3\Omega_bh^2$ & $N_{\rm eff}$ & $\tau$ & $\Omega_k$\cr
PKL$^{\kappa}+\Omega_k$ &     0.024 &     0.0037 &     0.0040 &      0.068 &     0.0069 &     0.0028 &       0.22 &     0.0018 &       0.19 &       0.12 &     0.0048 &     0.0021 \cr
PKJ$^{\kappa}+\Omega_k$ &     0.036 &     0.0053 &     0.0054 &      0.093 &     0.0073 &     0.0035 &       0.23 &     0.0020 &       0.20 &       0.13 &     0.0049 &     0.0022 \cr
PKL$^{\kappa}+\Omega_k-\Omega_e$ &     0.021 &     0.0030 & --- &      0.047 &     0.0066 &     0.0028 &       0.22 &     0.0018 &       0.19 &       0.11 &     0.0048 &     0.0021 \cr
PKJ$^{\kappa}+\Omega_k-\Omega_e$ &     0.034 &     0.0046 & --- &      0.059 &     0.0071 &     0.0034 &       0.23 &     0.0019 &       0.19 &       0.12 &     0.0049 &     0.0022 \cr
PKL$^g+\Omega_k$ &     0.063 &      0.016 &     0.0056 &      0.078 &     0.0073 &     0.0041 &       0.23 &     0.0020 &       0.20 &       0.14 &     0.0049 &     0.0016 \cr
PKJ$^g+\Omega_k$ &      0.11 &      0.027 &     0.0064 &       0.12 &     0.0076 &     0.0058 &       0.23 &     0.0023 &       0.22 &       0.20 &     0.0049 &     0.0021 \cr
PKL$^g+\Omega_k-\Omega_e$ &     0.059 &      0.016 & --- &      0.074 &     0.0072 &     0.0040 &       0.23 &     0.0020 &       0.20 &       0.14 &     0.0049 &     0.0015 \cr
PKJ$^g+\Omega_k-\Omega_e$ &     0.098 &      0.025 & --- &       0.10 &     0.0073 &     0.0052 &       0.23 &     0.0023 &       0.21 &       0.17 &     0.0049 &     0.0021 \cr
\hline
\hline
\end{tabular}
\caption{Same as Table~\ref{table:withomk}, but only for combinations of CMB with weak
  lensing power spectrum measurements, and CMB with galaxy power spectrum measurements.}
\label{table:cmbgalorcmbwl}
\end{center}
\end{table*}

\begin{table*}[th]\footnotesize
\begin{center}
\begin{tabular}{lcccccccccccccc|c}
\hline
\hline
Probe & $w_0$ & $\Omega_{d0}$ & $\Omega_e$ & $\sum{m_{\nu}} \rm{(eV)}$ & $n_s$ & ${dn_s \over d\ln k}$ & $10^{10}\Delta_R^2$ & $\Omega_ch^2$ & $10^3\Omega_bh^2$ & $N_{\rm eff}$ & $\tau$ & $\Omega_k$\cr
PKL$^{\kappa}$L$^g+\Omega_k$ &     0.015 &     0.0030 &     0.0033 &      0.054 &     0.0062 &     0.0024 &       0.23 &     0.0016 &       0.21 &       0.11 &     0.0051 &     0.0015 \cr
PKJ$^{\kappa}$J$^g+\Omega_k$ &     0.022 &     0.0045 &     0.0047 &      0.071 &     0.0073 &     0.0031 &       0.24 &     0.0020 &       0.24 &       0.13 &     0.0053 &     0.0021 \cr
PKL$^{\kappa}$L$^g$L$^s+\Omega_k$ &     0.012 &     0.0025 &     0.0029 &      0.051 &     0.0059 &     0.0024 &       0.23 &     0.0016 &       0.20 &       0.11 &     0.0051 &     0.0014 \cr
PKJ$^{\kappa}$J$^g$J$^s+\Omega_k$ &     0.010 &     0.0027 &     0.0038 &      0.066 &     0.0069 &     0.0031 &       0.24 &     0.0019 &       0.23 &       0.12 &     0.0053 &     0.0020 \cr
PKL$^{\kappa}_{\ell<1000}$L$^g_{\ell<1000}+\Omega_k$ &     0.017 &     0.0037 &     0.0037 &      0.059 &     0.0066 &     0.0032 &       0.23 &     0.0017 &       0.20 &       0.13 &     0.0052 &     0.0015 \cr
PKJ$^{\kappa}_{\ell<1000}$J$^g_{\ell<1000}+\Omega_k$ &     0.027 &     0.0063 &     0.0053 &      0.082 &     0.0076 &     0.0048 &       0.24 &     0.0022 &       0.24 &       0.18 &     0.0053 &     0.0020 \cr
PKL$^{\kappa}_{\ell<1000}$L$^g_{\ell<1000}$L$^s+\Omega_k$ &     0.012 &     0.0029 &     0.0033 &      0.056 &     0.0064 &     0.0032 &       0.23 &     0.0017 &       0.20 &       0.13 &     0.0052 &     0.0014 \cr
PKJ$^{\kappa}_{\ell<1000}$J$^g_{\ell<1000}$J$^s+\Omega_k$ &     0.011 &     0.0030 &     0.0047 &      0.079 &     0.0074 &     0.0047 &       0.24 &     0.0022 &       0.23 &       0.18 &     0.0053 &     0.0020 \cr
\hline
\hline
PKL$^{\kappa}$L$^g+\Omega_k-\Omega_e$ &    0.0094 &     0.0023 & --- &      0.036 &     0.0048 &     0.0023 &       0.22 &     0.0016 &       0.19 &      0.091 &     0.0050 &     0.0014 \cr
PKJ$^{\kappa}$J$^g+\Omega_k-\Omega_e$ &     0.015 &     0.0037 & --- &      0.043 &     0.0066 &     0.0031 &       0.24 &     0.0020 &       0.23 &       0.11 &     0.0053 &     0.0021 \cr
PKL$^{\kappa}$L$^g$L$^s+\Omega_k-\Omega_e$ &    0.0081 &     0.0022 & --- &      0.036 &     0.0048 &     0.0023 &       0.22 &     0.0016 &       0.19 &      0.091 &     0.0050 &     0.0014 \cr
PKJ$^{\kappa}$J$^g$J$^s+\Omega_k-\Omega_e$ &    0.0081 &     0.0026 & --- &      0.042 &     0.0065 &     0.0031 &       0.24 &     0.0019 &       0.22 &       0.11 &     0.0053 &     0.0020 \cr
PKL$^{\kappa}_{\ell<1000}$L$^g_{\ell<1000}+\Omega_k-\Omega_e$ &     0.012 &     0.0034 & --- &      0.036 &     0.0052 &     0.0032 &       0.22 &     0.0017 &       0.18 &       0.10 &     0.0051 &     0.0014 \cr
PKJ$^{\kappa}_{\ell<1000}$J$^g_{\ell<1000}+\Omega_k-\Omega_e$ &     0.021 &     0.0059 & --- &      0.044 &     0.0067 &     0.0045 &       0.24 &     0.0022 &       0.22 &       0.14 &     0.0053 &     0.0020 \cr
PKL$^{\kappa}_{\ell<1000}$L$^g_{\ell<1000}$L$^s+\Omega_k-\Omega_e$ &    0.0092 &     0.0028 & --- &      0.036 &     0.0052 &     0.0032 &       0.22 &     0.0017 &       0.18 &       0.10 &     0.0051 &     0.0014 \cr
PKJ$^{\kappa}_{\ell<1000}$J$^g_{\ell<1000}$J$^s+\Omega_k-\Omega_e$ &    0.0084 &     0.0030 & --- &      0.043 &     0.0067 &     0.0044 &       0.24 &     0.0021 &       0.22 &       0.14 &     0.0053 &     0.0020 \cr
\hline
\hline
\end{tabular}
\caption{Same as Tables~\ref{table:withomk} and \ref{table:nosnova}, without any cross-correlations between the observables (except for tomographic ones within a survey).}
\label{table:nocross}
\end{center}
\end{table*}


We now explore each of these tables in greater
detail. Table~\ref{table:withoutomk} shows us that the dominant constraint on
the fraction of dark energy at early times is drawn from the CMB (in
particular $TT$ and to some extent $T\kappa_c$) due to its deep
redshift information. At a value of $\sigma(\Omega_e) = 8.6 \times
10^{-3}$ (Table~\ref{table:withoutomk}), the Planck CMB temperature and polarization
constraint is within a percent of the critical density. In general, EDE is best constrained 
by the CMB, followed by weak lensing tomography, galaxy tomography and SNe in
that order. For comparison, the low-redshift lensing constraint from
LSST on $\Omega_e$ is a factor of four (factor of six for JDEM) weaker
than from the CMB. If we impose a nonlinear cutoff to the convergence
spectra at $\ell_{\rm max}$ = 1000, the situation becomes more dire,
as the LSST and JDEM lensing constraints become worse by another
factor of three. Similarly, the galaxy tomography constraint from LSST
on $\Omega_e$ is a factor of five (factor of nine for JDEM) weaker
than the CMB constraint, and the LSST SN constraint is a factor of 70
weaker (factor of 30 for JDEM) than the CMB.  

Nevertheless, once the six observables ($T$, $E$, $\kappa_c$,
$\kappa$, $g$, $s$) and all relevant cross-correlations (see
Eqns.~\ref{eq:fisher}-\ref{eq:clall}) from Planck (or EPIC) and LSST
(or JDEM) are analyzed in a combined setting, the constraint on
$\Omega_e$ improves by a factor of four over the CMB
constraint.
The combined constraints are equally strong 
regardless of the choice of LSST or JDEM for the non-CMB observations
($\kappa$, $g$, $s$). For a JDEM-like experiment, the cross-correlations
improve the $\Omega_e$ constraints by a factor of about 2. 

As expected, we find the late-redshift parameters more strongly
constrained by the non-CMB probes. 
For example, in a universe where we
allow for the existence of early dark energy, the LSST weak lensing
constraints on the present DE density ($\Omega_{d0}$) and EOS ($w_0$) of
1\% and 6\% are much better than the constraints obtained from just
CMB lensed data of about 7\% and 20\% on present DE density and EOS
(Table \ref{table:withoutomk}).  
Galaxy tomography measurements with LSST
constrain $\Omega_{d0}$ and $w_0$ to 5\% and 10\%, respectively,
whereas the strongest SN constraints are derived from JDEM, at 10\%
and 5\% for $\Omega_{d0}$ and $w_0$ respectively. When we combine the
probes of lensing and galaxy tomography, SNe, and CMB, the parameter
constraints improve by a factor of seven in $w_0$ and factor of four
in $\Omega_{d0}$ compared to the constraints from the strongest single
probe, here weak lensing from LSST. 

The results of the joint analysis don't change significantly when we
relax the assumption of spatial flatness. 
The exception to this statement is $\Omega_{d0}$ for JDEM, which
degrades by about a factor of 2 for the case where EDE density is
fixed (see Table~\ref{table:withoutomkandede}). This is because for JDEM 
$\Omega_{d0}$ is most strongly constrained by SNe measurements,
which require a tight bound on the curvature.
In the joint analysis, the curvature density is constrained to $~6
\times 10^{-4}$ of the critical density, which is an order of
magnitude stronger than solely with the CMB temperature and
lensing. The ability to measure curvature down to this level is an
exciting possibility that has been highlighted previously
\cite{Knox:2005hx}. Our constraints on the curvature in the joint
analysis, and from combining the CMB exclusively with galaxies or weak
lensing in Fig.~\ref{table:cmbgalorcmbwl}, are consistent with the
results in Refs.~\cite{Knox:2005hx,Knox:2006ux}, {\it even with the
  introduction of early dark energy.}

\begin{figure}[!t]
\begin{center}
\vspace{-1.0em}
\centerline{\includegraphics[scale=0.46]{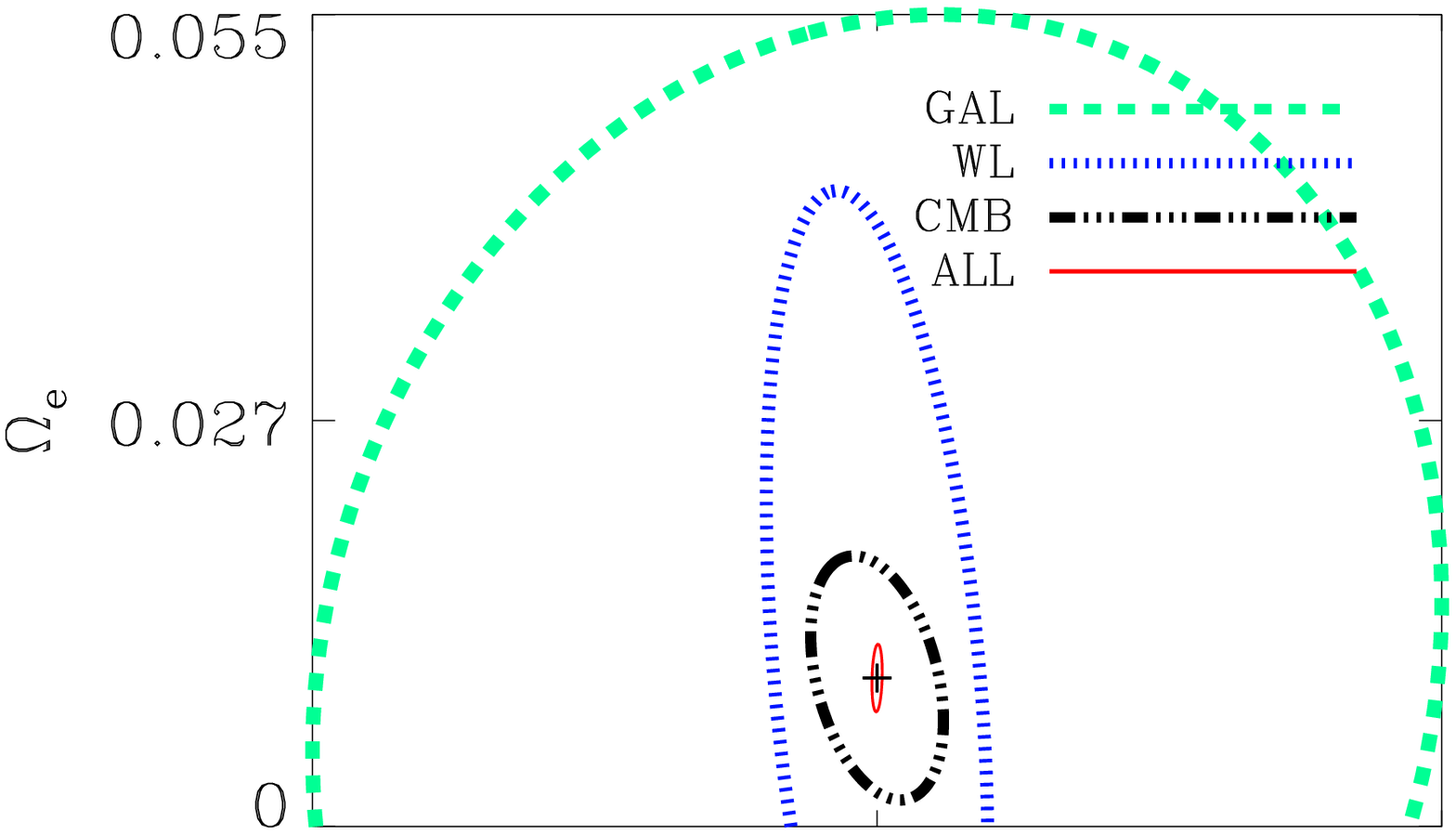}}
\vspace{-4.4em}
\centerline{\includegraphics[scale=0.46]{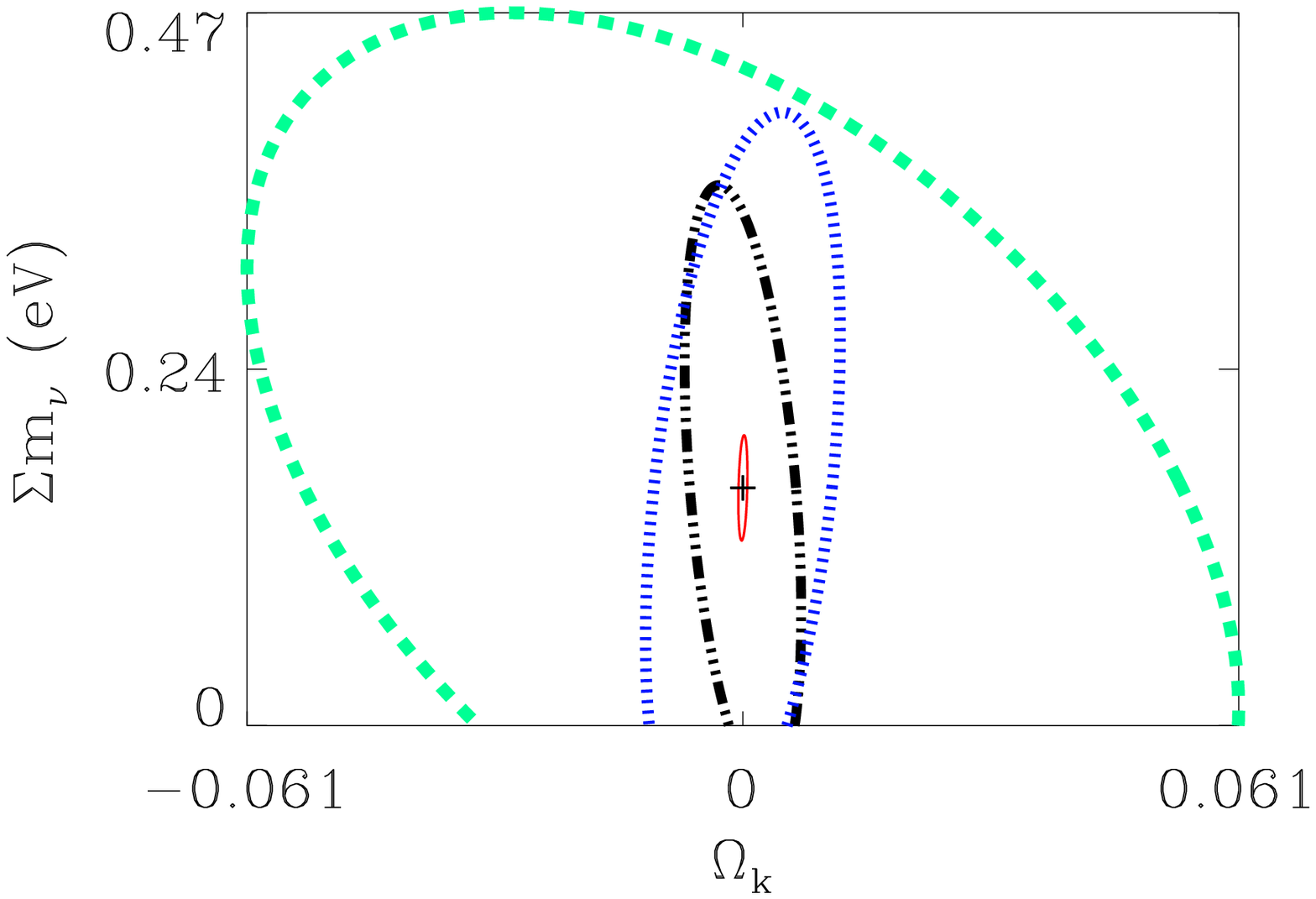}}
\end{center}
\vspace{-3em}
\caption{Error ellipses showing degeneracies between curvature
  ($\Omega_k$), early dark energy density ($\Omega_e$) and sum of
  neutrino masses ($\sum{m_\nu})$. The curves have the 
  same meaning as in Fig.~\ref{fig:ellipses}. Note the strong
  degeneracy between curvature and sum of neutrino masses for CMB data
  (black, dot-dashed) that was pointed out in Ref.~\cite{Smith:2006nk}. This
  degeneracy is broken when information on curvature from measures of
  cosmological distances are included, as shown by the solid (red)
  contour. The constraints in the plane of $\Omega_e$ and $\sum{m_\nu}$ are shown in
  Fig.~\ref{fig:ellipses}.}   
\label{fig:omkellipses}
\end{figure}

\begin{figure}[!t]
\begin{center}
\vspace{-0.6em}
\includegraphics[width=3.4in]{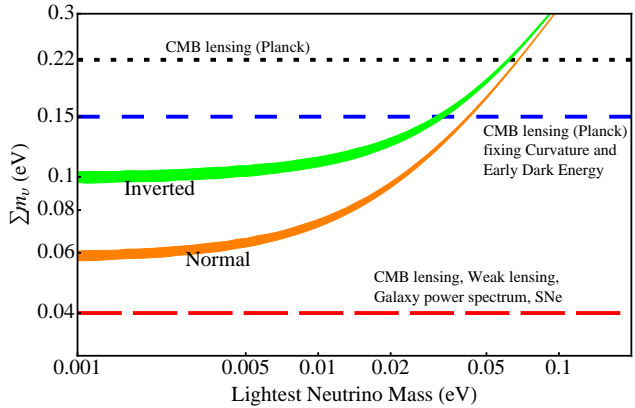}
\end{center}
\vspace{-1.7em}
\caption{Bands show the allowed values for the sum of neutrino masses as a function
  of the lightest neutrino mass eigenstate. The branch with higher
  values is for the inverted mass hierarchy. We have used the
  constraints on the neutrino mass squared differences from the
  Particle Data Group~\cite{Nakamura:2010zzi} to create bands of the
  allowed regions. The 3 horizontal lines show the 1$\sigma$ Fisher
  matrix error estimates for the cases ${\rm PK}+\Omega_k$ (0.22 eV, 
  Table~\ref{table:withomk}), ${\rm PK}-\Omega_e$ (0.15 eV,
  Table~\ref{table:withoutomkandede}) and ${\rm PKL}^\kappa {\rm L}^g{\rm
    L}^s+\Omega_k$ (0.04 eV, Table~\ref{table:withomk}).}
\label{fig:numixing}
\end{figure}

The sum of neutrino masses is most strongly constrained by the CMB
temperature and lensing spectra (including their cross-correlation), at the level
of 0.20 eV when the curvature is fixed (Table~\ref{table:withoutomk}). When 
curvature is allowed to vary, the sum of neutrino mass constraint degrades to 0.22 eV. 
There is a strong degeneracy between sum of neutrino masses and
curvature as pointed out in Ref.~\cite{Smith:2006nk}, and this is evident
in the bottom panel of Fig.~\ref{fig:omkellipses}. However, in the joint
analysis the constraint on curvature (of $6 \times 10^{-4}$ of the present critical density) is strong
enough to break this degeneracy, which will allow for an
exquisite measurement of the sum of neutrino masses. 
Our forecasts including CMB lensing, weak lensing tomography, galaxy
tomography, and supernovae show that constraints on the sum of neutrino 
masses at the level of 0.03-0.04 eV is possible (see Tables~\ref{table:withoutomk} 
and \ref{table:withomk}). At this level, a detection is possible if the neutrino 
mass hierarchy is inverted as Fig.~\ref{fig:numixing} shows.  

When the early dark energy density and curvature are both fixed, the
constraint on sum of neutrino masses is 0.15 eV from Planck alone. This
clearly shows that the sum of neutrino masses is substantially
correlated with early dark energy. This is also apparent in the right
column top panel of Fig.~\ref{fig:ellipses} where we plot the
constraints in the plane of the sum of neutrino masses and EDE density. Thus, as
expected, ignorance of the expansion history at high redshifts can
significantly hamper the constraints on the sum of neutrino masses from CMB 
lensing and cosmic shear. However, the constraints (even in the
presence of EDE) can be substantially improved by adding cosmic shear
tomography data because the degeneracy direction in cosmic shear 
data is substantially different from that in CMB lensing (see
Fig.~\ref{fig:ellipses}).   

We note here that our estimate of 0.15 eV when curvature and EDE
density are held fixed is based on the assumption that reionization
happened sharply, and therefore all of the effects on the CMB can be
encapsulated in the optical depth parameter. We also allow for more
complicated reionization histories by imposing a floor of $\ell=30$ in
the polarization data and a prior on $\tau$ of
0.01~\cite{Kaplinghat:2003bh}. With this setup, we find that neutrino
mass constraints degrade by about 10\%  (see
${\rm P}_{\ell>30}{\rm K}+\sigma(\tau)$ cases in
Tables~\ref{table:withoutomk},\ref{table:withomk},\ref{table:withoutomkandede}). 

We also considered a case assuming sharp reionization where $w_0$ is
held fixed as is EDE density and curvature. In this rather optimistic
setup, we obtain an uncertainty of about 0.14 eV for the sum of neutrino
masses with Planck lensing. We note that this is about 20\% less
stringent than the number quoted in Ref.~\cite{dePutter:2009kn}, some
of which can be traced to their different fiducial cosmology and
larger sky fraction (75\% vs 65\%).

We checked that the cross-correlation of $\kappa_c$ and $T$
(i.e. without $\kappa_c\kappa_c$) doesn't particularly improve the
constraint on the sum of neutrino masses, but it improves the
constraint on the late time dark energy density and EOS by about 50\%
compared to temperature and polarization data alone. 

Compared to the CMB with lensing, the LSST constraint on the sum of
neutrino masses is moderately weaker for both galaxy tomography (by
$50\%$) and weak lensing tomography (by $25\%$), whereas the JDEM
constraint on the sum of neutrino masses is a factor of three weaker
for galaxy tomography and  a factor of two weaker for weak lensing
tomography. In analyzing all the probes allowing for EDE density to
vary, we find a one-sigma constraint of 0.03-0.04 eV for LSST and JDEM  
(see Table \ref{table:withoutomk}). Tables~\ref{table:nosnova}
and~\ref{table:nocross} show that the strength of JDEM in this joint
analysis is owed partly to the superior SN constraints on the dark
energy EOS (which helps in breaking degeneracies) and partly to the
cross-correlations between the temperature, lensing, and galaxy
fields. 

\begin{figure}[!t]
\epsfxsize=3.4in
\vspace{-1.2em}
\centerline{\epsfbox{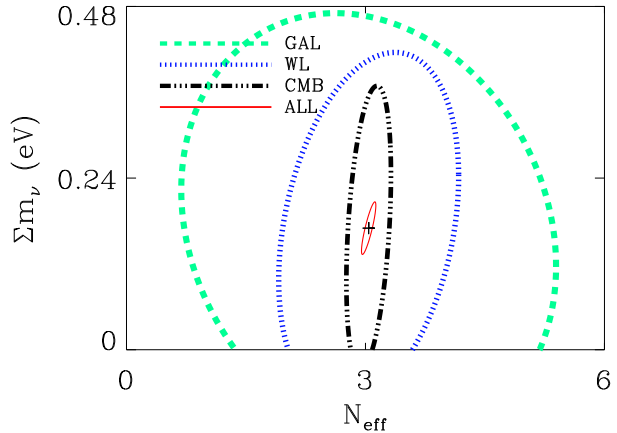}}
\vspace{-1.0em}
\caption{Error ellipses showing degeneracies between the sum of neutrino masses
  ($\sum{m_{\nu}}$) and the effective number of neutrinos ($N_{\rm{eff}}$) in a flat universe, where we have marginalized over the early dark energy density. The curves have the 
  same meaning as in Fig.~\ref{fig:ellipses}. Even for our extended parameter spaces, the CMB temperature as measured by Planck will be able to determine the possible existence of extra relativistic species, 
  while the combination of experimental probes is extremely helpful in pinning down the sum of neutrino masses.}   
  \label{fig:numassneff}
\vspace{-0.2em}
\end{figure}

Given the recent $2\sigma$ indications of extra relativistic species in the universe (e.g.~$N_{\rm eff} = 3.86 \pm 0.42$ with the datasets WMAP7+SPT+HST+BAO for a minimal cosmology~\cite{Keisler:2011aw}; for extended parameter spaces see~\cite{Joudaki:2012fx}), our understanding of $N_{\rm eff}$ depends critically on the ability to distinguish its signatures from other cosmological parameters. 
Fortunately, even for our extended parameter spaces, Planck alone could determine the possible existence of extra relativistic species in the universe at the $4\sigma$ level, as seen in Tables~\ref{table:withoutomkandede} and \ref{table:withomk}.
The constraints on $N_{\rm eff}$ are only weakly correlated with the constraints on $\sum{m_{\nu}}$, as shown in Fig.~\ref{fig:numassneff}.
When Planck is combined with JDEM or LSST, this allows for a more than 10$\sigma$ detection of additional light degrees of freedom in our universe.

Turning our attention to systematics, Table~\ref{table:withoutomk} shows that
the different choices for 
nonlinear cutoffs in $\ell_{\rm max} \in [1000, 3000]$ do not 
significantly affect the combined constraints.
We can therefore throw out highly
nonlinear scales ($\ell > 1000$) without significant loss in the
parameter constraints shown in this work. 
This is primarily due to the combination of cosmological probes and to
a lesser extent, due to the cross-correlations we have introduced. 
For comparison, considering only one probe, the
constraint degradation could be up to a factor of 3 in dark
energy and neutrino mass parameters (Table~\ref{table:withoutomk}). 
This insensitivity to nonlinear scales is maintained even without SN
measurements (Table~\ref{table:nosnova}). 

Comparing Table~\ref{table:withoutomk} with Table~\ref{table:withomk}, i.e.~comparing 
constraints for a flat universe with a universe that allows for a possible non-zero curvature, shows
the CMB temperature, polarization, and lensing constraints on $\Omega_k$ from Planck 
improve by an order of magnitude when accounting for future weak lensing, galaxy 
clustering, and supernova measurements from LSST or JDEM.
The SN measurements are, as expected, most sensitive to the curvature
prior, exhibiting significant degradations in parameters across the board.
This motivates the consistent use of a CMB curvature prior for SN measurements. 
The parameter that the SN observations with a JDEM-like survey
measure well is the dark energy EOS (the relative weakness of SNe
from LSST is explained in Sec.~\ref{snlabel}). 
Tables~\ref{table:withomk} and~\ref{table:nosnova} show that the neglect of
SN observations has $<10\%$ impact on the cosmological constraints for
LSST, whereas for a JDEM survey the dark energy parameters degrade by
up to 50\% without SN measurements. The same features are found in the
scenarios with an $\ell_{\rm max}=1000$ nonlinear cutoff in multipole space.  

\begin{table*}[th]\footnotesize
\begin{center}
\begin{tabular}{lcccccccccccccc|c}
\hline
\hline
Probe & $w_0$ & $\Omega_{d0}$ & $\Omega_e$ & $\sum{m_{\nu}} \rm{(eV)}$ & $n_s$ & ${dn_s \over d\ln k}$ & $10^{10}\Delta_R^2$ & $\Omega_ch^2$ & $10^3\Omega_bh^2$ & $N_{\rm eff}$ & $\tau$ & $\Omega_k$\cr
${\rm P}+\Omega_k$ &      0.21 &      0.042 &     0.0041 &       0.32 &     0.0042 &     0.0049 &       0.15 &     0.0017 &      0.082 &       0.14 &     0.0031 &     0.0069 \cr
${\rm PK}+\Omega_k$ &     0.092 &      0.024 &     0.0038 &      0.079 &     0.0040 &     0.0044 &       0.14 &     0.0013 &      0.077 &       0.11 &     0.0029 &     0.0024 \cr
PKL$^{\kappa}$L$^g$L$^s+\Omega_k$ &    0.0062 &     0.0015 &     0.0017 &      0.027 &     0.0035 &     0.0015 &       0.13 &    0.00074 &      0.064 &      0.052 &     0.0027 &    0.00046 \cr
PKJ$^{\kappa}$J$^g$J$^s+\Omega_k$ &    0.0065 &     0.0019 &     0.0020 &      0.029 &     0.0037 &     0.0019 &       0.13 &    0.00088 &      0.066 &      0.058 &     0.0028 &    0.00061 \cr
PKL$^{\kappa}_{\ell<1000}$L$^g_{\ell<1000}$L$^s+\Omega_k$ &    0.0065 &     0.0016 &     0.0018 &      0.028 &     0.0036 &     0.0024 &       0.13 &    0.00076 &      0.066 &      0.061 &     0.0028 &    0.00051 \cr
PKJ$^{\kappa}_{\ell<1000}$J$^g_{\ell<1000}$J$^s+\Omega_k$ &    0.0067 &     0.0020 &     0.0021 &      0.030 &     0.0038 &     0.0029 &       0.13 &    0.00097 &      0.067 &      0.069 &     0.0028 &    0.00065 \cr
\hline
\hline
\end{tabular}
\caption{Same as Table~\ref{table:withomk}, except for a 2m EPIC survey in lieu of Planck.}
\label{table:epic}
\end{center}
\vspace{-0.4em}
\end{table*}

\begin{table*}[th]\footnotesize
\begin{center}
\begin{tabular}{lccccccccccccc|c}
\hline
\hline
Probe & $w_0$ & $\Omega_{d0}$ & $\Omega_e$ & $\sum{m_{\nu}}$ & $n_s$ & ${dn_s \over d\ln k}$ & $10^{10}\Delta_R^2$ & $\Omega_ch^2$ & $10^3\Omega_bh^2$ & $N_{\rm eff}$ & $\tau$ \cr
PKL$^{\kappa}$L$^g-\Omega_e$ &      0.83 &       0.45 & --- &       0.23 &       0.41 &      0.091 &       0.37 &       0.40 &      0.061 &       0.41 &       0.29 \cr
PKJ$^{\kappa}$J$^g-\Omega_e$ &       2.4 &        1.5 & --- &       0.17 &       0.35 &       0.27 &       0.73 &       0.34 &     0.029 &       0.39 &       0.48 \cr
\hline
\hline
\end{tabular}
\caption{We present the fractional bias ${{|\delta(p_i)|} / {\sigma(p_i)}}$ (see Eqn.~\ref{biaseqn}) on cosmological parameters for the full combination of lensing, galaxy, and CMB surveys up to $\ell=2000$. The letters [L, J] in the first column denote [LSST, JDEM] (precise explanation of notation in caption of Table~\ref{table:withoutomk}). The parameter biases are comparable for Planck and EPIC. In this table, we assume a flat fiducial universe with $\Omega_e = 0.01$, and determine the offset in the parameter estimates for the case where EDE is not accounted for in the analysis. Naturally, the bias is larger for a universe with a larger fraction of EDE.}
\label{table:bias}
\end{center}
\end{table*}

Tables~\ref{table:withomk} and~\ref{table:nocross} show that most
parameters improve by $20\%-40\%$ when cross-correlations are included
for LSST. The cross-correlations have a greater
impact for JDEM than for LSST. In particular, the early dark energy
density, baryon and CDM densities, and sum of neutrino mass constraints improve by
up to factor of 2 with cross-correlations, the curvature density constraint improves by factor 
of 3 (also for LSST), while the improvement in the other
parameters are at the same $20\%-40\%$ level as for LSST. 
Naturally, when we weaken the parameter constraints by imposing
$\ell_{\rm max} = 1000$ in galaxy and lensing tomography, the impact
of the cross-correlations increases somewhat (e.g. an additional $20\%$ in early dark energy 
and sum of neutrino masses for JDEM). These quantitative comparisons between a
full covariance and one without cross-correlations largely hold true
independently of the SN sample. In the final analysis with all
cross-correlations, both JDEM and LSST are expected to provide similar
parameter constraints even though the individual constraints and
systematics are different.   

We now turn attention to the impact of keeping EDE fixed. In
Table~\ref{table:withoutomkandede} we explore the case where EDE exists but is kept
fixed for the Fisher matrix analysis. We expect this to mimic the
analysis where the existence of EDE is neglected. We explore biases
resulting from such a scenario in the ensuing section. 
We expect the constraints in this limit (of keeping
$\Omega_e$ fixed) to match the constraints in a non-EDE cosmology,
i.e., the standard case with $\Omega_e=0$.
Table~\ref{table:withoutomkandede} shows that the CMB temperature and
polarization constraints on cosmology are nearly unaffected
($\lsim10\%$ differences) by the removal of EDE. The removal of 
EDE improves the sum of neutrino masses by 30\%, and most other
parameter constraints by $\lsim 10\%$ when CMB lensing is included.

If we turn to the non-CMB probes, we find that significant degeneracies
with $\Omega_e$ exist for the dark energy parameters, sum of neutrino masses,
and the parameters determining the normalization, spectral index, and
its running of the matter power spectrum, as expected from
Fig.~\ref{fig:pk}. For instance, for weak lensing with both LSST and 
JDEM, the power spectrum normalization constraint improves by more
than a factor of two without EDE, the present DE EOS improves by 70\%,
the running by 50\%, the sum of neutrino masses and spectral index by 30\%, and
the present DE density by 20\%. 
When we throw out nonlinear scales ($\ell_{\rm max} = 1000$), the
constraint on the sum of neutrino masses only improves by about 5\%.
This is because the sum of neutrino masses derives its
degeneracy with the EDE density from small angular scales, as also
illustrated by the respective parameter derivatives in
Fig.~\ref{fig:wlbins}. 

In SN measurements, we find the strongest
degeneracy between the present dark energy density and the early dark
energy density, which degrades the constraint on the present DE
density by factor of five in LSST and factor of seven in JDEM. In the
joint observational analysis of prospective cosmological constraints,
we also find that our ignorance of EDE significantly degrades our 
constraints on the other cosmological parameters. As an example, when 
keeping $\Omega_e$ fixed in Table~\ref{table:withoutomkandede}, the DE EOS
constraint of $\sigma(w_0) \simeq 3 \times 10^{-3}$ (for both LSST and
JDEM) has improved by a factor of 3 for LSST and factor of 2 for JDEM.
In Table~\ref{table:nosnova}, we find these degradations would have
been stronger if SN observations were not considered in the analysis. 
We further note that this 0.3\% constraint on the EOS 
may be translated into roughly 10 model-independent redshift bins each constrained at 
the 1\% level (since $\sum_{i} {\sigma^{-2}(w_i)} = {\sigma^{-2}({w})}$~\cite{Joudaki}, where $w_i$ are binned EOS parameters and ${w}$ refers to a constant EOS).
Thus, if the dark energy EOS varies at late times, there is significant promise in detecting its variation with our considered surveys.

To end this section, we consider the constraints achievable with a
future CMB mission such as the 2m EPIC (EPIC-2m)
proposal~\cite{Baumann:2008aq, Bock:2009xw} in  
Table~\ref{table:epic}.
The properties of this probe are listed in Table~\ref{table:cmbspecs}. With
up to $40\%$ increased resolution and factor of thirty lower noise in an individual 
band than Planck, the EPIC-2m survey in Table~\ref{table:epic} shows
improved constraints across the board compared to Planck. 
The joint observational constraints, with EPIC-2m in lieu of Planck,
also show significant gains -- about $30\%$ improvement in the DE 
constraints, about 40\% improvement in the sum of neutrino masses constraint, 
and up to a factor of two improvement in the other parameters.

\subsection{Dark Energy Bias}

We have analyzed the expected constraints on early dark energy from
use of weak lensing tomography, galaxy tomography, SNe, and CMB. If we
live in a universe with EDE but do not account for this in the
analysis of the data sets of these future probes, we will compute
cosmological parameter constraints that are not only overly
optimistic, but the best estimates of the parameters will
also be shifted.  

We estimate here the parameter bias that would arise from
assuming a fiducial universe with $\Omega_e = 0.01$, but where EDE is
not accounted for in the constraint analysis. The bias in each
parameter is given by~\cite{Knoxetal, Shapiro:2008yk, LoVerde:2006cj,
  Rybicki}: 
\begin{equation}
\delta p_{\alpha} = F^{-1}_{\alpha \beta} \sum_{l, \beta} \Delta\ell \times  
\Tr \left [ {{\bf \tilde C}}_\ell^{-1} {{\partial {\bf
C}_\ell} \over {\partial p_\beta}} {{\bf \tilde C}}_\ell^{-1} \delta C_\ell \right ],
\label{biaseqn}
\end{equation}
where $\delta C_\ell$ accounts for the difference between the convergence
spectra in a universe with EDE and a universe without EDE. 

Table~\ref{table:bias} presents the cosmological parameter biases
for the LSST and JDEM surveys combined with Planck (comparable biases
when combined with EPIC). For LSST, we find less than $50\%$ bias in
most parameters except for $w_0$ (about 80\%). Biases when
including JDEM instead of LSST are close to a factor
of 2 times the expected constraints for the present DE density and
EOS. We note here that these estimates for the biases assume a
fiducial cosmology with $\Omega_e=0.01$. The biases would be larger in
a  universe with a larger fraction of EDE. If we are to avoid
significant bias in future joint analyses of cosmological parameters,
we must account for the uncertain high redshift nature of the dark
energy component. 

\section{Conclusions}

It is conceivable that in order to resolve the nature of the force
behind the late-time acceleration of the universe, we need a deeper
understanding of how the universal expansion and growth of structure
was affected by the dark energy at earlier times.  
We have chosen an approach that combines probes with varying degrees
of sensitivity in different parts of cosmological parameter space, and
we have included the cross-correlations between different probes. We
have performed a joint Fisher matrix analysis of prospective
measurements of weak lensing tomography, galaxy 
tomography, SNe, and Planck CMB. 

Our results show that the best possible constraints on 
[$w_0$, $\Omega_{d0}$, $\Omega_e$, $\sum{m_{\nu}} \rm{(eV)}$, $n_s$,
${dn_s \over d\ln k}$, $10^{10}\Delta_R^2$, $\Omega_ch^2$,
$10^3\Omega_bh^2$, $N_{\rm eff}$, $\tau$, $\Omega_k$] will be at the
level of [0.0085, 0.0019, 0.0023, 0.038, 0.0051, 0.0021, 0.20, 0.0013, 0.13, 0.090, 0.0044, 0.00056] for a wide survey like LSST, and 
[0.0076, 0.0021, 0.0025, 0.036, 0.0055, 0.0024, 0.19, 0.0012, 0.14, 0.090, 0.0045, 0.00075] for a deep survey like JDEM.
These constraint assume that photometric redshift and other
systematic uncertainties in weak lensing can be controlled to better
than $0.1\%$.  Naturally, these constraints improve in a universe
without EDE. For constraints on our extended parameter space with present cosmological datasets, see Ref.~\cite{Joudaki:2012fx}.

Our main findings are summarized below.
\begin{itemize}
\item The dominant constraint on the fraction of dark energy at early
  times comes from the CMB (in particular $TT$ and to some extent
  $T\kappa_c$).
Constraints from Planck are expected to be within a percent of the
critical density. The next best probe of EDE is weak lensing
tomography, followed by galaxy tomography, and then SNe. 
When all 6 observables (CMB temperature $T$, CMB polarization $E$, CMB lensing
$\kappa_c$, weak lensing shear $\kappa$, galaxy counts $g$, SNe $s$)
and all relevant cross-correlations (see Eqns.~\ref{eq:fisher}-\ref{eq:clall}) 
are analyzed in a combined setting, the constraint on $\Omega_e$
improves by a factor of four over the CMB constraint. Interestingly,
due to the cross-correlations, the combined constraints are
equally strong for LSST and JDEM ($\kappa$, $g$, $s$).     

\item When analyzed together with early dark energy, we find that the
  sum of active neutrino masses is most strongly constrained by the
  CMB lensing potential power spectrum, at the level of 0.22 eV for
  Planck (0.20 eV when flatness assumed) and 0.08 eV for 
  Epic (0.07 eV when flatness assumed). A combined analysis of Planck, and LSST
  or JDEM, shows that future constraints at the level of 0.04 eV are 
  possible. The CMB lensing constraints improve by $30\%$, 
  and the combined constraints by $60\%$, when EDE is not
  allowed to vary. Our results suggest that these constraints are not
  significantly affected by our ignorance of the reionization history,
  but more detailed work on this issue is necessary. 

\item The additional degree of freedom from the early dark energy
  density degrades our ability to measure late-time dark energy. We
  find that the present DE density can be measured to 0.2\% of the
  critical density and and equation of state to about 0.01, which is a
  factor of roughly 2 and 3 degradation, respectively, in the
  constraints compared to the case when EDE is fixed. 

\item Our analysis suggests that throwing out nonlinear
  scales ($\ell > 1000$) may not result in significant degradation in
  future parameter measurements when multiple cosmological probes are
  combined. Including cross-correlations improves parameter
  constraints on dark energy density and sum of neutrino masses by up
  to a factor of 2 when these nonlinear scales are not included. 

\item The curvature of the universe can be constrained to
  $6\times10^{-3}$ of the critical density  from CMB temperature and
  lensing alone, and improved by an order of magnitude in the joint
  analysis in agreement with the results of
  Ref.~\cite{Knox:2006ux}. Measurement of the curvature of the
  universe with Planck will be good enough that weak lensing
  tomography, galaxy tomography, and supernova measurements will not
  be limited by our ignorance of the curvature.

\item Even a modest 1$\%$  of the critical density at high redshift in
  dark energy, if not accounted for, shifts cosmological parameters by
  1-2 $\sigma$. Therefore, it is crucial for measurements of the
  underlying cosmology that we avoid prejudices about the energy content
  of the high redshift universe.

\end{itemize}

We have shown that degeneracies between cosmological parameters, in
particular between early dark energy, curvature and sum of neutrino
masses, can be effectively broken by a joint analysis of weak lensing
tomography, galaxy tomography, SNe, and the CMB (temperature,
polarization, lensing). Our analysis included the 
cross-correlations between these different probes. We find that 
simultaneous measurements of dark energy density at present and at high redshift
with a precision of 0.2\% of the respective critical densities,
present dark energy equation of state with a precision of 0.01,  
curvature with a precision of 0.06\% of the present critical density,
and sum of neutrino masses with a precision of 0.04 eV are 
possible.

\smallskip
{\it Acknowledgements:} We thank Ujjaini Alam and James Bullock for
discussions at an early stage of this work. We also thank Alexandre
Amblard, Francesco De Bernardis, Gregory D. Martinez, Quinn Minor,
Paolo Serra, and Joseph Smidt for useful discussions. 
SJ acknowledges support from the US Dept.~of Education through GAANN 
at UCI. MK acknowledges support from NSF through grant 0855462 at UCI.

\end{document}